\documentclass[sigconf,screen]{acmart}

\AtBeginDocument{%
  }

\setcopyright{cc}
\setcctype{by}
\acmDOI{10.1145/3832783.3837486}
\acmYear{2026}
\copyrightyear{2026}
\acmISBN{979-8-4007-2882-2/2026/10}
\acmConference[ASE '26]{Proceedings of the 41st IEEE/ACM International Conference on Automated Software Engineering}{October 12--16, 2026}{Munich, Germany}
\acmBooktitle{Proceedings of the 41st IEEE/ACM International Conference on Automated Software Engineering (ASE '26), October 12--16, 2026, Munich, Germany}
\acmSubmissionID{ase26main-p1257-p}
\received{2026-03-26}
\received[accepted]{2026-06-18}

\usepackage{minted}
\usepackage{xspace}
\usepackage[utf8]{inputenc} % allow utf-8 input
\usepackage[T1]{fontenc}    % use 8-bit T1 fonts
\usepackage{hyperref}       % hyperlinks
\usepackage{url}            % simple URL typesetting
\usepackage{booktabs}       % professional-quality tables
\usepackage{amsfonts}       % blackboard math symbols
\usepackage{nicefrac}       % compact symbols for 1/2, etc.
\usepackage{microtype}      % microtypography
\usepackage{xcolor}         % colors
\usepackage[table]{xcolor}
\usepackage[dvipsnames]{xcolor}
\usepackage{soul}
\usepackage{enumitem}
\usepackage[most]{tcolorbox}
\usepackage[tikz]{bclogo}
\usepackage{wrapfig}
\usepackage[linesnumbered,ruled,vlined]{algorithm2e}
\usepackage{multirow}
\usepackage{multicol}
\usepackage{makecell}
\usepackage{rotating}
\usepackage{hhline}
\usepackage{tabularray}
\usepackage{booktabs}
\usepackage{arydshln}
\usepackage{hyperref}
\usepackage{tablefootnote}
\usepackage{pifont}
\usepackage{changepage}

\tcbuselibrary{listingsutf8}

\SetKwInput{KwInputs}{Inputs}                % Set the Input
\SetKwInput{KwOutputs}{Outputs}              % set the Output
\SetKwInput{KwOutput}{Output}              % set the Output

\newcommand{\approach}{\textsc{ReCodeAgent}\xspace}
\newcommand{\alphatrans}{\textsc{AlphaTrans}\xspace}
\newcommand{\oxidizer}{\textsc{Oxidizer}\xspace}

\newcommand{\skel}{\textsc{Skel}\xspace}
\newcommand{\syzygy}{\textsc{Syzygy}\xspace}
\newcommand{\crust}{\textsc{Crust}\xspace}

\newcommand{\revision}[1]{\textcolor{black}{#1}}

\newtcolorbox{monotextbox}{
    colback=yellow!10,
    colframe=black,
    boxrule=0.5pt,
    arc=2pt,
    left=4pt,
    right=4pt,
    top=2pt,
    bottom=2pt,
    fontupper=\ttfamily\footnotesize,
    enhanced,
}

\definecolor{problemblue}{RGB}{100,134,158}
\definecolor{idiomsgreen}{RGB}{0,162,0}
\definecolor{exercisebgblue}{rgb}{0,  .69,  .941}
\definecolor{deepgreen}{rgb}{0.0, 0.5, 0.0}
\definecolor{codegreen}{rgb}{0,0.6,0}
\definecolor{codegray}{rgb}{0.5,0.5,0.5}
\definecolor{codepurple}{rgb}{0.58,0,0.82}
\definecolor{backcolour}{rgb}{0.95,0.95,0.92}
\definecolor{redColor}{RGB}{255,0,0}
\definecolor{Gray}{gray}{0.1}
\definecolor{bubblegum}{rgb}{0.99, 0.76, 0.8}
\definecolor{cambridgeblue}{rgb}{0.64, 0.76, 0.68}
\definecolor{babypink}{rgb}{1.0, 0.82, 0.86}
\definecolor{lightcoral}{rgb}{0.94, 0.66, 0.66}
\definecolor{mistyrose}{rgb}{1.0, 0.89, 0.88}
\definecolor{orchidpink}{rgb}{0.95, 0.74, 0.80}
\definecolor{carnationpink}{rgb}{1.0, 0.65, 0.79}
\definecolor{ashgray}{rgb}{0.70, 0.75, 0.71}
\definecolor{celadon}{rgb}{0.67, 0.88, 0.69}
\definecolor{powderblue}{rgb}{0.69, 0.88, 0.90}
\definecolor{etonblue}{rgb}{0.59, 0.78, 0.64}
\definecolor{teagreen}{rgb}{0.82, 0.94, 0.75}

\newcommand{\killpunct}[1]{}

\begin{document}

\title{\approach: A Multi-agent Workflow for Language-Agnostic Translation and Validation of Large-Scale Repositories}

\author{Ali Reza Ibrahimzada}
\correspondingauthor
\orcid{0000-0002-3797-818X}
\affiliation{%
  \institution{University of Illinois Urbana-Champaign}
  \city{Urbana}
  \state{IL}
  \country{USA}
}
\email{alirezai@illinois.edu}

\author{Brandon Paulsen}
\orcid{0000-0001-7790-6570}
\affiliation{%
  \institution{Amazon}
  \city{Arlington}
  \state{VA}
  \country{USA}
}
\email{bpaulse@amazon.com}

\author{Daniel Kroening}
\orcid{0000-0002-6681-5283}
\affiliation{%
  \institution{Amazon}
  \city{Seattle}
  \state{WA}
  \country{USA}
}
\email{dkr@amazon.co.uk}

\author{Reyhaneh Jabbarvand}
\orcid{0000-0002-0668-8526}
\affiliation{%
  \institution{University of Illinois Urbana-Champaign}
  \city{Urbana}
  \state{IL}
  \country{USA}
}
\email{reyhaneh@illinois.edu}

% \renewcommand{\shortauthors}{Ibrahimzada et al.}

\begin{abstract}
Most repository-level code translation and validation techniques have been evaluated on a single source-target programming language (PL) pair, owing to the complex engineering effort required to adapt new PL pairs. \emph{Programming agents} can enable PL-agnosticism in repository-level code translation and validation: they can synthesize code across many PLs and autonomously use existing tools specific to each PL's analysis. However, state-of-the-art has yet to offer a fully autonomous agentic approach for \emph{repository-level} code translation and validation of \emph{large-scale} programs. This paper proposes \approach, an autonomous multi-agent approach for language-agnostic repository-level code translation and validation. Users only need to provide the project in the source PL and specify the target PL for \approach to automatically translate and validate the entire repository. \approach is the \emph{first} technique to achieve high translation success rates across many PLs.

We compare the effectiveness of \approach with four alternative neuro-symbolic and agentic approaches to translate $118$ real-world projects, with $1{,}975$ LoC and $43$ translation units for each project, on average. The projects cover $6$ PLs and $4$ PL pairs. Our results demonstrate that \approach consistently outperforms prior techniques on translation correctness, improving test pass rate by $60.8\%$ on ground-truth tests, with an average cost of $\$15.3$. We also perform process-centric analysis of \approach trajectories to confirm its procedural efficiency. Finally, we investigate how the design choices (a multi-agent vs. single-agent architecture) influence \approach performance: on average, the test pass rate drops by $40.4\%$, and trajectories become $28\%$ longer and persistently inefficient.
\end{abstract}




\maketitle

\section{Introduction}
\label{sec:introduction}

Repository-level code translation---the process of converting an entire codebase from one programming language (PL) to another---is critical to improving software reliability and security and minimizing technical debt~\cite{jamshidi2013cloud,jain2015modernization,khadka2014professionals,nisar2022modernization}. Early work developed rule-based approaches~\cite{c2rust,c2go,sharpen,java2csharp}, like C2Rust, where all translation rules are written by hand. Later work developed neuro-symbolic techniques~\cite {cai2025rustmap,luo2025integrating,nitin2026c2saferrust,wang2026evoc2rust,zhou2026sactor,dehghan2025translating,ibrahimzada2025alphatrans,zhang2025oxidizer,shetty2024syzygy,wang2025skel,wang2026effireasontrans,yuan2026project}, which combine large language models (LLMs) with (symbolic) program analysis and testing. More recently, agentic approaches have been evaluated~\cite{khatry2025crust,guan2025repotransagent,li2025adversarial,sim2025large}, wherein one or more LLM agents work together to translate code between specific source-target PLs.

Prior rule-based and neuro-symbolic translation techniques only evaluate on a single source-target PL pair~\cite{c2rust,ibrahimzada2025alphatrans,wang2025skel,zhang2025oxidizer,nitin2026c2saferrust,shetty2024syzygy,dehghan2025translating,xue2025classeval,yang2024exploring}. This is due to the enormous engineering effort required to support a PL as a source or target language. The implementation of rule-based tools can go over $100K$ LoC\footnote{C2Rust~\cite{c2rust} (100K+ LoC).} and neuro-symbolic tools over 10K LoC\footnote{\alphatrans~\cite{ibrahimzada2025alphatrans} ($11K$ LoC), \oxidizer~\cite{zhang2025oxidizer} ($19K$ LoC), and \skel~\cite{wang2025skel} ($4K$ LoC).} just to support a single source-target PL pair. Given the quadratic number of PL pairs, scaling these techniques to many PL pairs is impractical. A PL-agnostic approach can help translate and validate projects across multiple PL-pairs without the need for complex engineering and external third-party dependencies.
% \reyhan{a missing motivation here is that, why do we even want a PL-agnostic pipeline? Please take a crack to justify. highlight the newly added text so that I can easily make a pass.}

Theoretically, agents can enable PL-agnostic code translation. However, having an end-to-end agentic code translation and validation pipeline can be challenging due to the following limitations:

\begin{enumerate}[wide, label=\textbf{(\arabic*)}]

    \item \textbf{True PL-agnosticism.} Existing agentic scaffolds operate on iterative reasoning-action-observation principle~\cite{yao2022react}.
    % \reyhan{cite ReAct paper}
    The actions performed through tool use are one of the key components that enable agent autonomy. In the context of code translation and validation, the tools can help agents explore and analyze the codebase, determine translation units, and validate translations. Existing agentic code translation techniques either employ basic, naive tools to only explore the codebase~\cite{khatry2025crust} or use PL-specific tools~\cite{li2025adversarial}. \revision{For example, \crust's SWE-agent uses only file navigation and build/test commands and underperforms a generate-then-repair loop~\cite{khatry2025crust}; ACToR depends on C/Rust CLI differential fuzzing~\cite{li2025adversarial}; and RepoTransAgent's tools are hardcoded for Java--C\# lookups~\cite{guan2025repotransagent}.}
    % \reyhan{when explaining the approach after challenges, mention that use of MCP, which is a universal semantic interface for any (or a lot of?) PLs enable PL-agnosticism in \approach} \ali{Added an initial version later in introduction.}

    \item \textbf{Hallucination in Repository-level Code Translation and Validation.} Translating large-scale repositories with tens or hundreds of files, specifically when translation and validation are integrated, is a long-horizon task~\cite{erdogan2025plan,chen2025reinforcement,sun2026scaling}, with hallucination being the main challenge for agents in this task. In the context of code translation, this includes hallucinating about class/file/method/variable names, finding matching libraries, or translating test assertions. 
    % \reyhan{polish up and may replace/add better examples}
    Without specific consideration of trajectory context and of hallucinations, an agentic code translation and validation may generate code that does not compile or preserve the original functionality. \revision{Long-horizon agents lose high-level goals without an explicit plan~\cite{erdogan2025plan}, confabulate under multi-step tool feedback~\cite{chen2025reinforcement}, and degrade as interaction history grows unbounded~\cite{sun2026scaling}; in repository translation, \alphatrans shows LLMs hallucinate types (including non-existent ones), method names, and assert APIs~\cite{ibrahimzada2025alphatrans}.}

    \item \textbf{Dichotomy of Test Translation.} Translating first and validating next approach simply does not work for large-scale repository-level translation because of long call chains and test coupling effect~\cite{ibrahimzada2025alphatrans}. \revision{In \alphatrans, each test averages $3$ direct callees and $27$ executed methods, so one buggy callee blocks validating the rest~\cite{ibrahimzada2025alphatrans}.} Such systems mostly use existing developer-written tests to validate functional equivalence, either through test translation and execution or language interoperability. Given that language interoperability may not exist for arbitrary PL pairs, a PL-agnostic approach may operate on test translation (of existing tests) and additional test generation. Code generation and validation, in general, are two conflicting objectives that should not be performed by one agent~\cite{lin2025learning,mcaleese_llm_2024,huang_agentcoder_2023,qian_chatdev_2024,dong_self-collaboration_2024,islam_mapcoder_2024}; \revision{e.g., co-generating code and tests biases the tests~\cite{huang_agentcoder_2023}.}
    %\reyhan{Ali, please add those 6 citations Brandon shared in the placeholder}
    Otherwise, the agent may modify the test rather than fix the incorrect code, e.g., by removing or relaxing assertions\revision{---\citet{ahmed2025cure} reports $21.8\%$--$33.0\%$ passing tests generated but fail golden tests in SWE-bench}. At the same time, test translation in a real-world setting is known to be even more challenging than code translation~\cite{abid2024gluetest,pan2024lost}, requiring the code as context to understand the structure of complex objects. As a result, a naive separation of agents, one for translation and one for validation, may not work. 
    
    %This is because most frontier LLMs are fine-tuned via reinforcement learning to maximize a single reward function; thereby, in the presence of conflicting objectives, they may hijack one~\cite{karwowski2023goodhart}. 
    %Due to conflicting objectives, integration of translation and validation is not trivial in an agentic setting.%~\cite{pan2024feedback,pan2024spontaneous,liu2024llms,li2024llms,roytburg2026llm}.
    % \reyhan{Ali, please try to find some research papers that back up this hypothesis. Also, make a reference to the ablation study} \ali{added some citations. it mostly goes by "bias in LLM evaluation and in-context reward hacking."}

    \item \textbf{Transparent and Process-centric Evaluation.} Prior agentic techniques rarely discuss the principled design space of an agentic code translation workflow~\cite{li2025adversarial,guan2025repotransagent,khatry2025crust}. They do not evaluate how architectural choices influence the final translation quality. Although they all validate translations through test execution, there's a dearth of discussion on the limitations of test translation~\cite{ibrahimzada2025alphatrans}, e.g., deletion of assert statements during test translation, generation of new tests with none to low-quality assertions, and threats to the validity of findings due to corresponding false positives~\cite{ke2025advancing}. There is also a dearth of transparency in cost reporting, and no attempts to analyze trajectories beyond final translation outcomes. \revision{Outcome-centric metrics alone are insufficient: on \crust, SWE-agent reaches the same test pass rate as a generate-repair loop while wasting many steps on file navigation~\cite{khatry2025crust}; ACToR and RepoTransAgent report primarily pass/compile rates~\cite{li2025adversarial,guan2025repotransagent}. Test outcomes can also mislead: \alphatrans notes that automatically generated tests can raise pass rates with weak assertions~\cite{ibrahimzada2025alphatrans}, TRAM measures $2.4\%$/$7.6\%$ false positives/negatives under GraalVM-based validation~\cite{ke2025advancing}.}
    
\end{enumerate}

% \reyhan{When explaining the approach, specifically mention that how your architecture resolved those challenges (analyzer for PL-agnoxticism, plannar for hallucination, and separation of task logics and inner loop for the conflicting objective challenge.} \ali{made a first attempt to do this. will polish more.}

\begin{figure*}[t]
    \centering
    \includegraphics[width=0.85\textwidth]{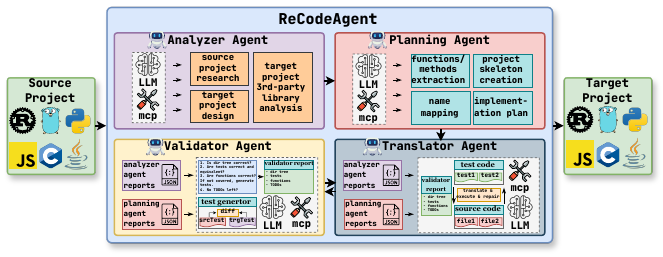}
    \vspace{-15pt}
    \caption{Overview of \approach.}
    \label{fig:overview}
    \vspace{-10pt}
\end{figure*}

This paper presents \approach, a multi-agent framework for \emph{language-agnostic}, repository-level code translation and validation (\S\ref{sec:approach}). \approach leverages four specialized agents, dividing the overall task into distinct phases: analysis, planning, translation, and validation. The Analyzer Agent (\S\ref{subsec:analyzer-agent}) explores the source project to create a high-level translation design and determine idiomatic alternatives of third-party libraries in the target PL using  
%using PL-agnostic static analysis exposed as 
Model Context Protocol (MCP) tools to the agent.
%design for the project's translation (\S\ref{subsec:analyzer-agent}). It first researches about the source project and overcomes the challenge of translation context and prompt engineering, previously raised by RepoTransAgent~\cite{guan2025repotransagent} and other neuro-symbolic techniques~\cite{ibrahimzada2025alphatrans,zhang2025oxidizer,wang2025skel}. 
The Planning Agent (\S\ref{subsec:planning-agent}) identifies translation units, constructs a concrete project skeleton, and devises a plan with specific steps for subsequent agents.
%to follow for translation and validation. 
This agent specifically addresses the complex engineering effort required by existing neuro-symbolic techniques---\emph{challenge 1}---by replacing their PL-dependent program analysis component (e.g., dependency graph construction using CodeQL~\cite{codeql}) with tool-assisted, LLM-centered static analysis (lightweight tools that support many PLs, e.g., Tree-sitter~\cite{tree-sitter}). 
%It uses lightweight analysis tools that support many PLs, e.g., Tree-sitter~\cite{tree-sitter} and Language Server Protocols (LSPs)~\cite{agrawal_monitor-guided_2023}. 
% \reyhan{shall we mention lcp tools here as well?} \ali{that would be a bit unnecessary!} \reyhan{but isn't how you justify pl-agnosticism through their semantic interface?}. 
The plan guides subsequent agents with concrete steps to avoid unnecessary exploration, which can cause hallucinations (\emph{challenge 2}).
The Translator agent (\S\ref{subsec:translator-agent}) carries out the steps outlined in the plan to co-translate code and tests. The Validator agent (\S\ref{subsec:validator-agent}) executes the translated tests or generates new tests to validate the translation. Separating the dynamic validation workflow from the Validator agent, while the Translator agent translates both code and test, addresses \emph{challenge 3}. Translating tests is essential for the pipeline's generalizability beyond command-line tools, compared with \citet{guan2025repotransagent} and \citet{li2025adversarial}. 
%Translation and validation occur in two distinct agents to ensure non-conflicting objectives and bias in LLM agents.
%Unlike existing agentic techniques~\cite{guan2025repotransagent,li2025adversarial}, the translator and validator agents can translate both source and test code, enabling \approach to translate projects beyond command line tools compared to \citet{li2025adversarial}. 
% To enhance the effectiveness of each agent, \approach incorporates the language server protocol as a tool, enabling reliable, real-time access to codebase information.

We evaluate the effectiveness of \approach 
%for repository-level code translation and validation 
against four existing neuro-symbolic and agentic techniques~\cite{zhang2025oxidizer,wang2025skel,ibrahimzada2025alphatrans,khatry2025crust}. Our benchmark comprises $4{,}583$ translation units, drawn from $118$ real-world projects totaling over $230K$ LoC (\S\ref{subsec:experiment-setup}). Prior techniques translate these projects between four PL pairs, namely C-Rust, Go-Rust, Java-Python, and Python-JavaScript.
%which cover $4$ PL pairs, drawn from $118$ real-world projects totaling over $230K$ lines of code (\S\ref{subsec:experiment-setup}). 
%For each benchmark, we translate and validate all source and test files using \approach. 
On average, the translated projects by \approach are $99.4\%$ and $86.5\%$ correct in terms of compilation success and test pass rate, $2.5\%$ and $60.8\%$ higher than those of alternative approaches (\S\ref{subsec:rq1}). When translating tests, \approach achieves $99.3\%$, $0.91$, and $94.9\%$ assertion equivalence, cosine similarity, and assertion type match, respectively, demonstrating their quality
%its ability in correctly translating tests in target PL 
(\S\ref{subsec:rq2}). Ablation study shows that removing the Analyzer, Planning, and Validator agents reduces the test pass rate by $22.7\%$, $25.3\%$, and $30.3\%$, respectively, while increasing trajectory complexity by $28\%$, measured by two process-centric metrics~\cite{liu2026process}. %\reyhan{what's this metric? trajectory steps or grapehctory metrics?} \ali{28\% is the increase in terms of both NC and TEC, both Graphectory metrics} \reyhan{if there are three metrics, then why only one number?} \ali{i took the average}
Comparison with two baseline agents indicates 
that they significantly underperform 
%that it performs significantly worse than 
\approach, achieving only $25.3\%$ ($\downarrow$$61.2\%$) and $24.1\%$ ($\downarrow$$62.4\%$) test pass rate (\S\ref{subsec:rq3}). \approach is 
%cheap to develop ($9{,}498$ LoC) and 
cost-effective, translating and validating projects in $57$ minutes with \revision{a} cost of $\$15.3$, on average (\S\ref{subsec:rq4}). These results confirm that \approach is a viable alternative to prior approaches and is vastly easier to adapt to new PL pairs. Our contributions are:

\begin{enumerate}[label=\textbf{(\arabic*)}, leftmargin=*]
    
    \item \emph{\textbf{Technique.}} \approach is the first multi-agent, PL-agnostic pipeline for repository-level code translation and validation. It does not require major engineering effort or dependency on external tools.

    \item \emph{\textbf{Empirical Evaluation.}} We rigorously evaluate \approach on $118$ real-world repository-level projects and $4$ PL pairs against the state-of-the-art neuro-symbolic and agentic techniques. The results indicate that \approach outperforms existing techniques in repository-level code translation and validation, without the need for PL-specific engineering effort.

    \item \emph{\textbf{Tool.}} The implementation of \approach, the agent logs and trajectories required to reproduce the results presented in this paper are publicly available~\cite{website}.

\end{enumerate}

\section{Problem Definition and Architecture Design}
\label{sec:overview}

% \begin{table*}[t]
%     % \setlength{\tabcolsep}{1pt}
%     \footnotesize
%     \centering
%     \caption{Details of MCP tools used in \approach.}
%     \vspace{-10pt}
%     \input{Resources/Tables/tools}
%     \label{table:tools}
%     \vspace{-10pt}
% \end{table*}

% A source project $P_s = (F_s, T_s, D_s)$ consists of source functions $F_s = \{f_1^s, \ldots, f_n^s\}$, tests $T_s$, and dependencies $D_s$, written in language $L_s$. Given a target language $L_t$, the \emph{repository-level code translation problem} is to produce $P_t = (F_t, T_t, D_t)$ such that: (1)~$P_t$ compiles without errors, (2)~$\forall i, \forall \vec{x}: \llbracket f_i^s \rrbracket(\vec{x}) \simeq \llbracket f_i^t \rrbracket(\vec{x})$, i.e., corresponding functions are semantically equivalent, and (3)~all translated tests pass. Here, $\vec{x}$ denotes function inputs, $\llbracket \cdot \rrbracket$ denotes semantic interpretation, and $\simeq$ denotes observational equivalence.

A source project $P_s = (F_s, T_s, D_s)$ consists of source functions $F_s = \{f_s^1, \ldots, f_s^n\}$, tests $T_s$, and dependencies $D_s$, written in language $L_s$. Given a target language $L_t$, the \emph{repository-level code translation problem} is to produce $P_t = (F_t, T_t, D_t)$ such that: (1)~$P_t$ compiles without errors, (2)~$\forall i, \forall \vec{x_s}, \forall \vec{x_t}: \vec{x_s} \mapsto \vec{x_t} \implies \llbracket f_s^i \rrbracket(\vec{x_s}) \simeq \llbracket f_t^i \rrbracket(\vec{x_t})$, i.e., corresponding functions are semantically equivalent, and (3)~all translated tests $T_t$ pass. Here, $\vec{x_s}$ and $\vec{x_t}$ denote function inputs in languages $L_s$ and $L_t$, respectively, $\mapsto$ is a mapping of concrete values in $L_s$ to $L_t$, $\llbracket \cdot \rrbracket$ denotes semantic interpretation, and $\simeq$ denotes observational equivalence. However, given the practical limitations of current testing and verification techniques, in practice we only consider a subset of all inputs to each $f_s^i$ and $f_t^i$.

Figure~\ref{fig:overview} presents an overview of \approach, which takes $P_s$ and $L_t$ as input and produces $P_t$. It consists of four components: the \emph{Analyzer Agent} (\S\ref{subsec:analyzer-agent}), \emph{Planning Agent} (\S\ref{subsec:planning-agent}), \emph{Translator Agent} (\S\ref{subsec:translator-agent}), and \emph{Validator Agent} (\S\ref{subsec:validator-agent}). The first two analyze $P_s$ and generate a dependency-aware implementation plan, while the last two execute an iterative translate--validate--repair loop to produce $P_t$.

The \emph{Analyzer Agent} performs extensive analysis of $P_s$. It analyzes the codebase and produces a report summarizing project structure, data models, classes, interfaces, structs, error-handling strategy, and $D_s$. It then analyzes library usage in $L_s$ by consulting documentation and identifying suitable counterparts in $L_t$. This component concludes by producing a target project design document that specifies how modules should be translated and which libraries should be used to preserve functionality.

The \emph{Planning Agent} decomposes the translation task into concrete sub-tasks: it identifies all functions in $F_s$ that require translation and constructs a consistent name mapping to ensure uniformity in $P_t$. It also generates a skeleton structure for $P_t$, outlining file organization and module boundaries, and produces an implementation plan with concrete tasks for translating and validating $P_s$.

The \emph{Translator Agent} and \emph{Validator Agent} execute the implementation plan. The \emph{Translator Agent} translates $F_s$ and $T_s$ into $F_t$ and $T_t$, incrementally filling in the skeleton files; if validation fails, it uses the Validator Agent's report to repair translation bugs. The \emph{Validator Agent} independently validates $P_t$ by executing $T_t$ and performing coverage-gap analysis; when functions in $F_t$ are uncovered, it triggers additional test generation and reports results back to the Translator Agent for repair.
\section{\approach}
\label{sec:approach}

Tools are essential and key components to enable autonomy for agents. In this section, we first explain the tools that assist \approach agents with static analysis (\S \ref{subsec:tools}) and explain the details and workflow of each agent for code translation and validation through reasoning and tool usage (\S \ref{subsec:analyzer-agent}--\S \ref{subsec:validator-agent}).

%followed by a discussion of each sub-agent.
%In this section, we provide a detailed explanation of each component of \approach, as shown in Algorithm~\ref{alg:approach}. We begin by explaining the tools that supply static analysis information, followed by discussing each sub-agent. We refer to Algorithm~\ref{alg:approach} when describing each sub-agent. For prompt templates used in \approach, we refer readers to our artifacts repository~\cite{website}.

\subsection{Tools}
\label{subsec:tools}

\begin{figure}[t]
  \centering
  \scriptsize
  % \vspace{-10pt}
  % First Image (The IDE Screenshot)
  \begin{minipage}{\linewidth}
    \centering
    \setlength{\fboxsep}{0pt}
    \fbox{\includegraphics[width=0.8\linewidth]{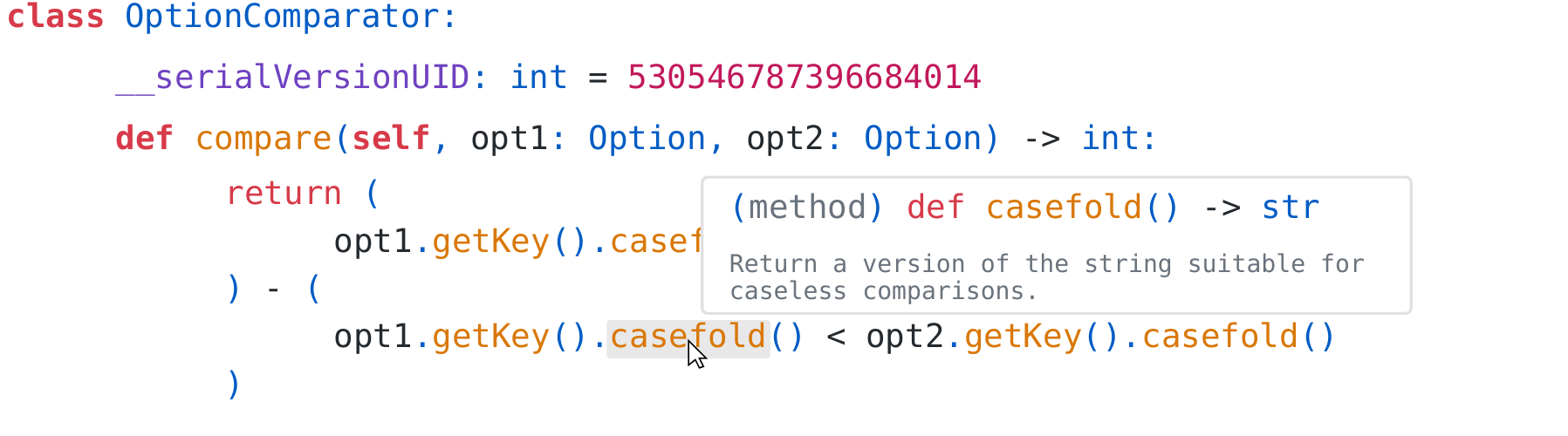}}
    \vspace{-10pt}
    \caption{Hover feature on a Python code in an IDE.
    % \reyhan{figure is blurry} \ali{its not that blurry when draft is downloaded as PDF. i'll try to fix it.}
    }
    \label{fig:hover-ide}
  \end{minipage}
  \hfill % Adds space between the two minipages
  % Second Image (The Code/Text Box)
  % \vspace{5pt}
  \begin{minipage}{\linewidth}
    \centering
    \tcbset{boxsep=2pt, top=2pt, bottom=2pt, before skip=0pt, after skip=0pt}
    \begin{monotextbox}
      \scriptsize
      \input{Resources/Figures/hover-2}
    \end{monotextbox}
    \vspace{-10pt}
    \caption{Hover tool from the Python LSP server.}
    \label{fig:hover-mcp}
  \end{minipage}
  \vspace{-15pt}
\end{figure}

\approach leverages a set of Model Context Protocol (MCP) tools to provide static analysis capabilities and facilitate effective agent interactions with the codebase. The MCP is an open-source protocol that allows LLMs to seamlessly connect to external data, tools, and software systems. We implement our custom tools and expose them to each agent in \approach through MCP. These tools enable agents to obtain detailed project information, modify code, and retrieve documentation in a PL-agnostic manner. 
% \reyhan{(1) Please briefly explain what MCP is for those who are not familiar. You may also use references. (2) why do you use mcp tools rather than you write your own tools? please explain and justify with references}
% Table~\ref{table:tools} summarizes the main MCP tools and their description, which are integrated into \approach:

\subsubsection{Language Server Protocol (LSP) Tools} LSP is an open, JSON-RPC-based protocol for use between source code editors or integrated development environments (IDEs) and servers that provide language intelligence: PL-specific features like code completion, syntax highlighting and marking of warnings and errors, as well as refactoring routines~\cite{agrawal_monitor-guided_2023}. The goal of the protocol is to allow PL support to be implemented and distributed independently of any given IDE. We use LSPs from six PLs~\cite{rustanalyzer,clangd,gopls,tslangserver,pylsp,jdtls} as a set of tools that allow agents to interact with the codebase at a semantic level, independent of the underlying PL. Extending support to more PLs only requires installing their language server which is usually maintained and available online~\cite{lspservers}, without writing any additional code.
% \reyhan{what's the overhead of supporting new PLs? please explain. Alternatively, if using them is as easy as updating a config file, mention that your approach supports x (x>20) PLs}
LSP functionalities that we \revision{integrate} include:

\begin{enumerate}[wide, label=(\arabic*), topsep=5pt]
    \item \texttt{definition}: It takes a symbol name as input and retrieves the complete implementation (e.g., function, class) along with the file path and line numbers from the codebase, enabling agents to understand and extract source code fragments for translation.
    
    \item \texttt{diagnostics}: Provides diagnostic information such as errors and warnings for a specified file, which assists agents in identifying potential issues in both source and translated code. IDEs usually indicate errors and warnings using red and yellow underlines in the code editor.
    
    \item \texttt{edit\_file}: Applies a set of text edits to a file atomically, supporting incremental construction and refinement of the translated project. This tool is helpful when there are a large number of edits that need to be applied, as opposed to using the agent's \texttt{\small Edit} tool once per every edit.
    
    \item \texttt{hover}: Returns documentation for a symbol at a specified position in the code, including docstrings and code deprecation details. Figure~\ref{fig:hover-ide} shows the hover feature in a normal IDE, while the same functionality from the LSP server is given in Figure~\ref{fig:hover-mcp}.
    
    \item \texttt{references}: Locates all occurrences and usages of a symbol across the codebase, essential for large-scale code refactoring by the agent. This tool returns exact line numbers of every usage, making it easy for agents to refer to them later.
    
    \item \texttt{rename\_symbol}: Renames a symbol at a given location and updates all corresponding references throughout the project. This tool is helpful in making consistent changes across the codebase, without the need to perform additional edits.
\end{enumerate}

\vspace{-5pt}
\subsubsection{Project Analysis (PA) Tools:} These tools extract structural information from the codebase, aiding agents in project comprehension and planning. The goal of these tools is to reduce the token consumption of the agent, which would otherwise be spent on exploring the codebase and files. Figure~\ref{fig:mcp-pa} shows two sample outputs from project analysis tools. Functionalities included are:

\begin{enumerate}[wide, label=(\arabic*)]
    \item \texttt{get\_directory\_tree}: Returns a structured representation of the project directory, printing files such as main, test, and configuration and their directory hierarchy.
    
    \item \texttt{get\_file\_structure}: Generates a structured representation of a given source file, identifying key code elements such as classes, functions, structs, and global variables.
\end{enumerate}

\begin{figure}[t]
  \centering
  \scriptsize
  % \vspace{-20pt}
  \begin{minipage}[t]{0.49\linewidth}
    \begin{monotextbox}
    \scriptsize
    \input{Resources/Figures/pa-1}
    \end{monotextbox}
  \end{minipage}
  \hfill
  \begin{minipage}[t]{0.49\linewidth}
    \begin{monotextbox}
    \scriptsize
    \input{Resources/Figures/pa-2}
    \end{monotextbox}
  \end{minipage}

  \vspace{-10pt}
  \caption{Project analysis (PA) tools output.}
  \label{fig:mcp-pa}
  \vspace{-15pt}
\end{figure}

\subsection{Analyzer Agent}
\label{subsec:analyzer-agent}

\begin{figure*}
    \centering

    \begin{minipage}[t]{0.325\textwidth}
        \begin{monotextbox}
        \scriptsize
        \input{Resources/Figures/analyzer-doc-1}
        \end{monotextbox}
    \end{minipage}
    \hfill
    \begin{minipage}[t]{0.325\textwidth}
        \begin{monotextbox}
        \scriptsize
        \input{Resources/Figures/analyzer-doc-2}
        \end{monotextbox}
    \end{minipage}
    \hfill
    \begin{minipage}[t]{0.325\textwidth}
        \begin{monotextbox}
        \scriptsize
        \input{Resources/Figures/analyzer-doc-3}
        \end{monotextbox}
    \end{minipage}

    \vspace{-10pt}
    \caption{Documents generated by \emph{Analyzer Agent} in \approach.}
    \label{fig:analyzer-agent-docs}
    \vspace{-10pt}
\end{figure*}

The \emph{Analyzer Agent} conducts initial research and formulates the high-level design of the translation (Algorithm~\ref{alg:approach}, line 22). This agent ensures that the target project preserves structurally similar to the source, and identifies the most suitable libraries and design patterns for the target PL. Figure~\ref{fig:analyzer-agent-docs} illustrates the three documents produced by this agent, corresponding to the following phases:

\subsubsection{Source Project Research}
\label{subsubsec:source-project-research}
The analyzer agent first explores the source codebase (e.g., using the \texttt{\small Read} tool) to ascertain its architectural design and functional requirements. Subsequently, it invokes the \texttt{\small get\_directory\_tree} tool to extract the project's directory structure, which serves as the foundational blueprint for translation. To get a semantic understanding of the codebase, the agent employs the \texttt{\small get\_file\_structure} and LSP tools to analyze the contents of each file in greater detail. The output of this phase is a research document that includes the source project's dependencies, error handling mechanisms, and directory hierarchy.

\subsubsection{Third-Party Library Analysis}
\label{subsubsec:library-analysis}
Next, the agent identifies all third-party and standard libraries utilized within the source project. For each identified dependency, the agent investigates idiomatic counterparts available in the target PL. Leveraging the \texttt{\small WebFetch} and \texttt{\small hover} tools, the agent retrieves official documentation to determine recommended usage patterns and evaluate the trade-offs associated with alternative library selections. These findings are consolidated into a document, ensuring that subsequent translation phases are guided by current best practices within the target PL ecosystem.

\subsubsection{Target Project Design}
\label{subsubsec:target-project-design}
In the final phase, the agent synthesizes its research into a comprehensive target project design document, which enforces a strict one-to-one structural mapping between source and target projects, covering directory structure, file organization, and identifier naming conventions (classes, methods, and variables). The document specifies which files require translation, how source constructs map to target equivalents (e.g., Java Interfaces $\rightarrow$ Rust Traits, Go Structs $\rightarrow$ Python Classes), and outlines strategies for error handling and library integration. This document serves as the authoritative reference for the subsequent \emph{Planning} (\S\ref{subsec:planning-agent}), \emph{Translator} (\S\ref{subsec:translator-agent}), and \emph{Validator} (\S\ref{subsec:validator-agent}) \emph{Agents}.

\begin{algorithm}[t]
    \caption{\small \approach}
    \label{alg:approach}    
    \scriptsize
    \SetKwInOut{Input}{Input}
\SetKwInOut{Output}{Output}

\SetKwProg{Fn}{Function}{:}{}

\SetKwData{sourceProject}{sourceProject}
\SetKwData{translatedProject}{translatedProject}
\SetKwData{translatedValidatedProject}{translatedValidatedProject}
\SetKwData{llm}{LLM}
\SetKwData{tools}{tools}
\SetKwData{timeout}{\textbf{\texttt{\scriptsize !timeout}}}
\SetKwData{timeoutt}{timeout}
\SetKwData{maxIteration}{maxIter}
\SetKwData{iteration}{iteration}
\SetKwData{in}{\textbf{\texttt{\scriptsize in}}}
\SetKwData{colon}{\textbf{\texttt{\scriptsize :}}}
\SetKwData{timeoutNotReached}{\timeout \colon}
\SetKwData{analyzerAgentClient}{analyzerAgent}
\SetKwData{planningAgentClient}{planningAgent}
\SetKwData{translatorAgentClient}{translatorAgent}
\SetKwData{validatorAgentClient}{validatorAgent}
\SetKwData{context}{context}
\SetKwData{analyzerAgentOutput}{analyzerOutput}
\SetKwData{planningAgentOutput}{planningOutput}
\SetKwData{implementationPlan}{implementationPlan}
\SetKwData{partA}{\texttt{\scriptsize Part-A}}
\SetKwData{partB}{\texttt{\scriptsize Part-B}}
\SetKwData{agentTranslatedSourceCode}{agentTranslatedSourceCode}
\SetKwData{agentTranslatedTestCode}{agentTranslatedTestCode}
\SetKwData{agentGeneratedTestCode}{agentGeneratedTestCode}
\SetKwData{validationReport}{validationReport}
\SetKwData{break}{\textbf{break}}
\SetKwData{plan}{plan}

\SetKwFunction{implement}{\textbf{implement}}
\SetKwFunction{repair}{\textbf{repair}}
\SetKwFunction{generateTests}{\textbf{generateAndValidateTests}}
\SetKwFunction{initializeAgent}{\textbf{initializeAgent}}
\SetKwFunction{analyzerAgent}{\textbf{runAnalyzerAgent}}
\SetKwFunction{planningAgent}{\textbf{runPlanningAgent}}
\SetKwFunction{translatorAgent}{\textbf{runTranslatorAgent}}
\SetKwFunction{validatorAgent}{\textbf{runValidatorAgent}}
\SetKwFunction{isAllSuccess}{\textbf{isAllSuccess}}
\SetKwFunction{validateTranslations}{\textbf{validateTranslations}}
\SetKwFunction{hasUncoveredFunctions}{\textbf{hasUncoveredFunctions}}

\Input{\sourceProject, \llm, \tools, \timeoutt $=5000s$, \maxIteration $=5$, \validationReport $\leftarrow$ $\varnothing$}
\Output{\translatedValidatedProject}

\Fn{\translatorAgent(\context, \validationReport)}{
    \translatorAgentClient $\leftarrow$ \initializeAgent(\llm, \tools, \context) \\

    \If{\validationReport $\neq$ $\varnothing$}{
        \Return{ \translatorAgentClient.\repair(\validationReport) }
    }
    \implementationPlan $\leftarrow$ \context.\planningAgentOutput.\plan \\ \translatedProject $\leftarrow$ $\varnothing$ \\
    \ForEach{\partA $\in$ \implementationPlan}{
        \agentTranslatedSourceCode $\leftarrow$ \translatorAgentClient.\implement(\partA) \\
        \translatedProject $\leftarrow$ \translatedProject $\cup$ \agentTranslatedSourceCode   
    }

    \ForEach{\partB $\in$ \implementationPlan}{
        \agentTranslatedTestCode $\leftarrow$ \translatorAgentClient.\implement(\partB) \\
        \translatedProject $\leftarrow$ \translatedProject $\cup$ \agentTranslatedTestCode
    }
    
    \Return{\translatedProject}
}

\Fn{\validatorAgent(\context, \translatedProject)}{
    \validatorAgentClient $\leftarrow$ \initializeAgent(\llm, \tools, \context) \\

    \validationReport $\leftarrow$ \validatorAgentClient.\validateTranslations(\translatedProject)

    \If{\validationReport.\hasUncoveredFunctions()}{
        \agentGeneratedTestCode $\leftarrow$ \validatorAgentClient.\generateTests() \\
        \translatedProject $\leftarrow$ \translatedProject $\cup$ \agentGeneratedTestCode
    }

    \translatedValidatedProject $\leftarrow$ \translatedProject \\

    \Return{\translatedValidatedProject, \validationReport}
}

\timeoutNotReached \analyzerAgentOutput $\leftarrow$ \analyzerAgent(\sourceProject) \\
\timeoutNotReached \planningAgentOutput $\leftarrow$ \planningAgent(\sourceProject, \analyzerAgentOutput) \\
\context $\leftarrow$ \sourceProject $\cup$ \analyzerAgentOutput $\cup$ \planningAgentOutput \\

\For{$\iteration \leftarrow 1$ \KwTo \maxIteration}{
    \timeoutNotReached \translatedProject $\leftarrow$ \translatorAgent(\context, \validationReport) \\
    \timeoutNotReached \translatedValidatedProject, \validationReport $\leftarrow$ \validatorAgent(\context, \translatedProject) \\

    \If{\validationReport.\isAllSuccess()}
    {
        \break
    }
}

\Return{\translatedValidatedProject}

\end{algorithm}

\subsection{Planning Agent}
\label{subsec:planning-agent}

\begin{figure*}
    \centering

    \begin{minipage}[t]{0.24\textwidth}
        \begin{monotextbox}
        \scriptsize
        \input{Resources/Figures/planning-doc-1}
        \end{monotextbox}
    \end{minipage}
    \hfill
    \begin{minipage}[t]{0.24\textwidth}
        \begin{monotextbox}
        \scriptsize
        \input{Resources/Figures/planning-doc-2}
        \end{monotextbox}
    \end{minipage}
    \hfill
    \begin{minipage}[t]{0.24\textwidth}
        \begin{monotextbox}
        \scriptsize
        \input{Resources/Figures/planning-doc-3}
        \end{monotextbox}
    \end{minipage}
    \hfill
    \begin{minipage}[t]{0.24\textwidth}
        \begin{monotextbox}
        \scriptsize
        \input{Resources/Figures/planning-doc-4}
        \end{monotextbox}
    \end{minipage}

    \vspace{-10pt}
    \caption{Documents generated by \emph{Planning Agent} in \approach.}
    \label{fig:planning-agent-docs}
    \vspace{-10pt}
\end{figure*}

The \emph{Planning Agent} reads the source project research and target project design documents generated by the \emph{Analyzer Agent} (\S\ref{subsec:analyzer-agent}) and decomposes the high-level design into granular, executable implementation steps (Algorithm~\ref{alg:approach}, line 23). This agent ensures that every source code file is translated and validated according to a logical, dependency-aware order. Figure~\ref{fig:planning-agent-docs} illustrates the documents and skeleton files produced by the planning agent.

\subsubsection{Fragment Extraction}
\label{subsubsec:fragment-extraction}
The agent first extracts all translation units---including functions, methods, and classes---from available files. Fragment extraction is performed using the \texttt{\small get\_file\_structure} tool, while maintaining a strict validation-in-the-loop process. For validation, the agent generates executable scripts to verify that every extracted fragment exists in the source codebase and that no files have been omitted. This validation step mitigates agent hallucination, wherein the agent erroneously concludes that a task has been completed when it has not. This phase concludes by generating a document of extracted fragments, with each fragment recorded in the format \texttt{\small file\_name:fragment\_name}.

\subsubsection{Name Mapping}
\label{subsubsec:name-mapping}
To ensure one-to-one translation and naming consistency, the agent constructs a mapping from source fragments to their target counterparts, strictly preserving symbol names (e.g., \texttt{\small camelCase}, \texttt{\small snake\_case}) to maintain functional parity across the project. This mapping is then used during skeleton generation to produce accurate method signatures in the target PL, preventing LLMs from arbitrarily renaming methods and classes in ways that impede translation tracking.

\subsubsection{Skeleton Generation}
\label{subsubsec:skeleton-generation}
The agent subsequently constructs the target project's directory structure and populates it with skeleton files. These skeleton files contain class declarations and method signatures without concrete implementations. This approach provides a compilable framework that mirrors the source project's architecture, enabling the translation process to proceed incrementally.

\subsubsection{Implementation Plan}
\label{subsubsec:implementation-plan}
The implementation plan is a structured document partitioned into source code translation (\texttt{\small Part A}) and test code translation and validation (\texttt{\small Part B}). The plan adheres to a bottom-up ordering, ensuring that dependencies are implemented prior to the modules that rely upon them. For example, a sample step in \texttt{\small Part A} can be \texttt{\small "Translate HelpFormatter.py and validate its syntactical correctness"}. Each step in the plan is expected to yield compilable code and provides an explicit checklist for the \emph{Translator} (\S\ref{subsec:translator-agent}) and \emph{Validator} (\S\ref{subsec:validator-agent}) \emph{Agents} to execute.

\subsection{Translator Agent}
\label{subsec:translator-agent}
The \emph{Translator Agent} carries out the implementation plan by executing both \texttt{\small Part A} (source code translation) and \texttt{\small Part B} (test translation) (Algorithm~\ref{alg:approach}, lines 1--13). The objective of this agent is to translate the source project into the target PL while preserving functional equivalence and architectural alignment. If the \emph{Validator Agent} reports failures, the Translator Agent enters repair mode and applies targeted fixes based on the validation report. The agent follows a systematic workflow to ensure a one-to-one translation:

\begin{enumerate}[wide, label=(\arabic*)]
    \item \textit{Context Integration:} The agent loads the implementation plan, the target design document, and the name mapping files. This ensures that translated identifiers (e.g., class and variable names) remain consistent with the plan and are not arbitrarily renamed.

    \item \textit{Incremental Implementation:} Following the dependency-aware ordering from the planning phase, the agent replaces stubs in the target skeleton files with complete implementations for \texttt{\small Part A}. It then translates developer-written tests for \texttt{\small Part B}, creating the corresponding test files in the target project.

    \item \textit{Language-Specific Adaptation:} When translating between languages with divergent feature sets, the agent applies targeted adaptation strategies that preserve behavior (e.g., emulating overloading via default arguments or dispatch).

    \item \textit{Repair Mode:} When provided with a non-empty validation report, the agent diagnoses the reported failures and updates the translated source and/or test code accordingly, iterating until validation succeeds or the iteration budget is exhausted.
\end{enumerate}

The output of this agent is a translated project that includes both translated source code and tests, ready for the \emph{Validator Agent}.

\subsection{Validator Agent}
\label{subsec:validator-agent}
The \emph{Validator Agent} validates the functional correctness of the translated project (Algorithm~\ref{alg:approach}, lines 14--21). Given a translated project (including translated tests) produced by the \emph{Translator Agent}, this agent executes tests, performs coverage-gap analysis, and produces a validation report that is fed back to the \emph{Translator Agent} for repair in the next iteration.

\subsubsection{Validation and Failure Reporting}
\label{subsubsec:test-translation}
The agent executes the translated test suite in the target environment and checks whether all tests pass. If failures occur (e.g., compilation errors or assertion failures), the agent consolidates diagnostics---including stack traces and failing test cases---into a structured validation report. This report identifies the failing functions and provides actionable feedback for the Translator Agent's repair step in the next iteration.

\subsubsection{Coverage-Guided Test Generation}
\label{subsubsec:test-generation}
To fully validate the translated modules, the agent performs a coverage-gap analysis by comparing the executed tests against the complete list of functions identified during the planning phase. If uncovered functions remain, it generates additional tests in both the source and target PLs to exercise the uncovered functions and adds them to the translated project. The generated tests are executed in both PLs to ensure matching behavior. The agent then re-executes validation to update the validation report, ensuring that the translated code is both functionally correct (with respect to the available tests) and more rigorously exercised. The iterative loop continues until all tests pass or the maximum iteration limit is reached.

\section{Evaluation}
\label{sec:evaluation}

To evaluate different aspects of \approach, we investigate the following research questions:

\begin{enumerate}[label=\bfseries RQ\arabic*:]

  \item \sloppy \textit{Effectiveness of \approach}. To what extent can \approach effectively translate real-world projects? Can it outperform expensively developed techniques?

  \item \textit{Test Translation}. To what degree are the translated tests equivalent to the original tests? What are the limitations of \approach when translating tests?

  \item \textit{Ablation Study}. To what extent do the Analyzer, Planning, and Validator agents impact the performance of \approach? Can a standalone LLM agent perform similarly to \approach?

  \item \textit{Cost and Tool Usage Analysis}. How much does it cost and how long does it take for \approach to translate projects? What kinds of tools are frequently invoked by \approach?

\end{enumerate}

\subsection{Experimental Setup}
\label{subsec:experiment-setup}

\subsubsection{Benchmark} We assess the performance of \approach using benchmarks from previously published studies on automated repository-level code translation and validation. Each benchmark contains a project implemented in a specific PL that includes both source and test code. The goal for each benchmark is to translate and validate the project in a target PL, ensuring that all tests are successfully executed and pass. Table~\ref{table:effectiveness} provides an overview of our open-source subject translation projects from four recent repository-level code translation techniques~\cite{khatry2025crust,ibrahimzada2025alphatrans,shetty2024syzygy,zhang2025oxidizer} covering the following PLs: C, Go, Rust, Java, Python, and JavaScript. \revision{The current implementation of \approach supports tools discussed in \S\ref{subsec:tools} for these six PLs. More languages can be supported by implementing \texttt{\small get\_file\_structure} with roughly $300$--$350$ LoC per PL. The effort usually involves creating the parse tree and traversing it to extract specific code constructs, e.g., functions, classes, variables, etc. LSP tools do not require manual effort, it can be simply integrated by downloading the PL's language server protocol implementation}. In total, our evaluation includes $118$ projects spanning over $230K$ lines of code. We exclude \syzygy~\cite{shetty2024syzygy} from our evaluation due to the unavailability of its artifact. Since the test suites of RepoTransBench~\cite{wang2025repotransbench} are written by LLMs and not validated as correct by humans, we exclude them as well. For \alphatrans~\cite{ibrahimzada2025alphatrans}, we select a subset of projects for which the authors have provided validated test suites. Moreover, the \crust benchmark consists of $100$ independent C projects translated to Rust. All these prior works produced translations of real-world open-source GitHub repositories and assessed functional equivalence via test execution.

\subsubsection{LLM} \approach works with different LLMs. Major software engineering leaderboards~\cite{swebench} have shown that the Claude Sonnet performs similarly to or in some cases outperforms other state-of-the-art proprietary LLMs, such as OpenAI GPT-5 and Google Gemini Pro. Therefore, we use Anthropic's \texttt{\small Claude 4.5 Sonnet} as the main LLM in all our experiments\footnote{\revision{Please see our artifacts~\cite{website} on how to run with other models from OpenRouter.}}. To make our results reproducible without re-running experiments, \approach logs the inputs, intermediate agent interactions, tool execution results, and outputs of the LLM, and supports replaying these logs. Each agent in \approach terminates within the budget of $5{,}000$ seconds, empirically set after analyzing the runtime of our largest project.

\subsubsection{Competing Techniques} We compare \approach against \skel~\cite{wang2025skel}, \oxidizer~\cite{zhang2025oxidizer}, \alphatrans~\cite{ibrahimzada2025alphatrans}, and SWE-agent~\cite{yang2024swe} from \crust~\cite{khatry2025crust} \revision{after reproducing their results with the same LLM}. While we cannot directly compare to ACToR~\cite{li2025adversarial} as its implementation is tied to CLI programs, our ablation that removes the analyzer and planner agents closely resembles its agent architecture, and can serve as a proxy for its performance\footnote{Confirmed by the authors of ACToR~\cite{li2025adversarial}.}.

\begin{table*}[t]
    % \setlength{\tabcolsep}{2pt}
    % \scriptsize
    \centering
    \caption{Effectiveness of \approach in repository-level code translation and validation in terms of test and function validation. \textbf{LoC:} Lines of Code, \textbf{CS:} Compilation Success, \textbf{TE:} \# Tests Executed, \textbf{TP:} \# Tests Passing, \textbf{TF:} \# Tests Failing, \textbf{CRUST-\{$\alpha,\beta,\sigma,\gamma$\}}: $\alpha$: both compile, $\beta$: only \approach compile, $\sigma$: only SWE-agent compile, $\gamma$: both do not compile, C: Test coverage, C+: \revision{Test coverage with generated tests}. Tuple entries indicate $\langle$Tool, \approach$\rangle$.}
    \vspace{-10pt}
    \resizebox{\textwidth}{!}{
        \begin{tabular}{c|c|c|c|c|c:c:c|c:c:c|c:c:c|c|c|c|c|c} 
\hline
\multirow{2}{*}{\begin{tabular}[c]{@{}c@{}}\textbf{Tool}\\\textbf{(PL Pair)}\end{tabular}}                                                       & \multirow{2}{*}{\textbf{Project}}      & \multirow{2}{*}{\textbf{LoC}} & \multirow{2}{*}{\textbf{CS~(\%)}} & \multirow{2}{*}[10pt]{\begin{tabular}[c]{@{}c@{}}\textbf{\# }\\\textbf{Validated}\\\textbf{Developer}\\\textbf{Tests}\end{tabular}} & \multicolumn{3}{c|}{\begin{tabular}[c]{@{}c@{}}\textbf{\textbf{Validated}}\\\textbf{\textbf{Developer Tests}}\end{tabular}}             & \multicolumn{3}{c|}{\begin{tabular}[c]{@{}c@{}}\textbf{\approach }\\\textbf{Translated~}\\\textbf{Developer Tests}\end{tabular}}        & \multicolumn{5}{c|}{\begin{tabular}[c]{@{}c@{}}\textbf{\approach }\\\textbf{Generated Tests}\end{tabular}}                                                                   & \multicolumn{3}{c}{\begin{tabular}[c]{@{}c@{}}\textbf{Function}\\\textbf{Validation}\end{tabular}}  \\ 
\cline{6-19}
                                                                                                                                                 &                                        &                               &                                   &                                                                                                                               & {\cellcolor[rgb]{1,1,0.792}}\textbf{TE} & {\cellcolor[rgb]{0.792,0.894,0.792}}\textbf{TP} & {\cellcolor[rgb]{1,0.792,0.792}}\textbf{TF} & {\cellcolor[rgb]{1,1,0.792}}\textbf{TE} & {\cellcolor[rgb]{0.792,0.894,0.792}}\textbf{TP} & {\cellcolor[rgb]{1,0.792,0.792}}\textbf{TF} & {\cellcolor[rgb]{1,1,0.792}}\textbf{TE} & {\cellcolor[rgb]{0.792,0.894,0.792}}\textbf{TP} & {\cellcolor[rgb]{1,0.792,0.792}}\textbf{TF} & \textbf{C (\%)} & \textbf{C+ (\%)} & \textbf{\textbf{Total}} & \textbf{\textbf{Success}}   & \textbf{Fail}                               \\ 
\hline
\multirow{6}{*}{\begin{tabular}[c]{@{}c@{}}\oxidizer\\(Go$\rightarrow$Rust)\end{tabular}}               & checkdigit           & 428                           & $\langle$100, 100$\rangle$        & 36                                                                                                                            & $\langle$36, 36$\rangle$                & $\langle$33, 36$\rangle$                        & $\langle$3, 0$\rangle$                      & 36                                      & 36                                              & 0                                           & 71                                      & 71                                              & 0                                           & 79.7            & 94.7             & 29                      & $\langle$21, 29$\rangle$    & $\langle$8, 0$\rangle$                      \\
                                                                                                                                                 & go-edlib                & 639                           & $\langle$100, 100$\rangle$        & 36                                                                                                                            & $\langle$36, 36$\rangle$                & $\langle$19, 36$\rangle$                        & $\langle$17, 0$\rangle$                     & 36                                      & 36                                              & 0                                           & 3                                       & 3                                               & 0                                           & 94.7            & 94.9             & 24                      & $\langle$18, 24$\rangle$    & $\langle$6, 0$\rangle$                      \\
                                                                                                                                                 & histogram           & 314                           & $\langle$100, 100$\rangle$        & 2                                                                                                                             & $\langle$2, 2$\rangle$                  & $\langle$2, 2$\rangle$                          & $\langle$0, 0$\rangle$                      & 2                                       & 2                                               & 0                                           & 66                                      & 66                                              & 0                                           & 38.0            & 90.5             & 19                      & $\langle$12, 19$\rangle$    & $\langle$7, 0$\rangle$                      \\
                                                                                                                                                 & nameparts           & 413                           & $\langle$100, 100$\rangle$        & 26                                                                                                                            & $\langle$26, 26$\rangle$                & $\langle$23, 26$\rangle$                        & $\langle$3, 0$\rangle$                      & 26                                      & 26                                              & 0                                           & 22                                      & 22                                              & 0                                           & 96.8            & 96.8             & 15                      & $\langle$9, 14$\rangle$     & $\langle$6, 1$\rangle$                      \\
                                                                                                                                                 & stats                     & 1241                          & $\langle$100, 100$\rangle$        & 121                                                                                                                           & $\langle$121, 121$\rangle$              & $\langle$71, 121$\rangle$                       & $\langle$50, 0$\rangle$                     & 121                                     & 121                                             & 0                                           & 320                                     & 320                                             & 0                                           & 43.3            & 79.6             & 52                      & $\langle$38, 52$\rangle$    & $\langle$14, 0$\rangle$                     \\
                                                                                                                                                 & textrank               & 1132                          & $\langle$100, 100$\rangle$        & 8                                                                                                                             & $\langle$8, 8$\rangle$                  & $\langle$6, 8$\rangle$                          & $\langle$2, 0$\rangle$                      & 8                                       & 8                                               & 0                                           & 127                                     & 127                                             & 0                                           & 72.6            & 98.7             & 52                      & $\langle$40, 52$\rangle$    & $\langle$12, 0$\rangle$                     \\ 
\hline
\rowcolor[rgb]{0.8,0.902,0.902} \textbf{Total}                                                                                                   &                                        & 4167                          & $\langle$100, 100$\rangle$        & 229                                                                                                                           & $\langle$229, 229$\rangle$              & $\langle$154, 229$\rangle$                      & $\langle$75, 0$\rangle$                     & 229                                     & 229                                             & 0                                           & 609                                     & 609                                             & 0                                           & 70.9            & 92.5             & 191                     & $\langle$138, 190$\rangle$  & $\langle$53, 1$\rangle$                     \\ 
\hline
\multirow{4}{*}{\begin{tabular}[c]{@{}c@{}}\alphatrans\\(Java$\rightarrow$Python)\end{tabular}} & cli                 & 37841                         & $\langle$100, 100$\rangle$        & 381                                                                                                                           & $\langle$66, 381$\rangle$               & $\langle$35, 360$\rangle$                       & $\langle$31, 21$\rangle$                    & 381                                     & 381                                             & 0                                           & 257                                     & 257                                             & 0                                           & 96.7            & 97.3             & 257                     & $\langle$196, 241$\rangle$  & $\langle$61, 16$\rangle$                    \\
                                                                                                                                                 & csv                 & 33072                         & $\langle$100, 100$\rangle$        & 298                                                                                                                           & $\langle$147, 298$\rangle$              & $\langle$3, 241$\rangle$                        & $\langle$144, 57$\rangle$                   & 298                                     & 298                                             & 0                                           & 192                                     & 190                                             & 2                                           & 84.4            & 85.8             & 213                     & $\langle$74, 211$\rangle$   & $\langle$139, 2$\rangle$                    \\
                                                                                                                                                 & fileupload   & 3567                          & $\langle$100, 100$\rangle$        & 39                                                                                                                            & $\langle$39, 39$\rangle$                & $\langle$36, 39$\rangle$                        & $\langle$3, 0$\rangle$                      & 39                                      & 39                                              & 0                                           & 208                                     & 208                                             & 0                                           & 38.7            & 71.6             & 25                      & $\langle$19, 25$\rangle$    & $\langle$6, 0$\rangle$                      \\
                                                                                                                                                 & validator     & 41605                         & $\langle$100, 100$\rangle$        & 463                                                                                                                           & $\langle$359, 463$\rangle$              & $\langle$114, 438$\rangle$                      & $\langle$245, 25$\rangle$                   & 463                                     & 435                                             & 28                                          & 132                                     & 131                                             & 1                                           & 65.0            & 73.0             & 409                     & $\langle$217, 397$\rangle$  & $\langle$192, 12$\rangle$                   \\ 
\hline
\rowcolor[rgb]{0.8,0.902,0.902} \textbf{Total}                                                                                                   &                                        & 116085                        & $\langle$100, 100$\rangle$        & 1181                                                                                                                          & $\langle$611, 1181$\rangle$             & $\langle$188, 1078$\rangle$                     & $\langle$423, 103$\rangle$                  & 1181                                    & 1153                                            & 28                                          & 789                                     & 786                                             & 3                                           & 71.2            & 81.9             & 904                     & $\langle$506, 874$\rangle$  & $\langle$398, 30$\rangle$                   \\ 
\hline
\multirow{8}{*}{\begin{tabular}[c]{@{}c@{}}\skel\\(Python$\rightarrow$JavaScript)\end{tabular}}              & bst                         & 123                           & $\langle$100, 100$\rangle$        & 11                                                                                                                            & $\langle$11, 11$\rangle$                & $\langle$11, 11$\rangle$                        & $\langle$0, 0$\rangle$                      & 11                                      & 11                                              & 0                                           & 6                                       & 6                                               & 0                                           & 89.7            & 99.0             & 21                      & $\langle$21, 21$\rangle$    & $\langle$0, 0$\rangle$                      \\
                                                                                                                                                 & colorsys               & 120                           & $\langle$100, 100$\rangle$        & 2                                                                                                                             & $\langle$2, 2$\rangle$                  & $\langle$2, 2$\rangle$                          & $\langle$0, 0$\rangle$                      & 2                                       & 2                                               & 0                                           & 46                                      & 46                                              & 0                                           & 87.0            & 91.3             & 9                       & $\langle$9, 9$\rangle$      & $\langle$0, 0$\rangle$                      \\
                                                                                                                                                 & heapq                     & 189                           & $\langle$100, 100$\rangle$        & 8                                                                                                                             & $\langle$8, 8$\rangle$                  & $\langle$7, 8$\rangle$                          & $\langle$1, 0$\rangle$                      & 8                                       & 8                                               & 0                                           & 11                                      & 11                                              & 0                                           & 91.6            & 91.6             & 24                      & $\langle$23, 24$\rangle$    & $\langle$1, 0$\rangle$                      \\
                                                                                                                                                 & html                       & 684                           & $\langle$100, 100$\rangle$        & 7                                                                                                                             & $\langle$7, 7$\rangle$                  & $\langle$6, 7$\rangle$                          & $\langle$1, 0$\rangle$                      & 7                                       & 7                                               & 0                                           & 13                                      & 13                                              & 0                                           & 77.1            & 86.1             & 42                      & $\langle$39, 42$\rangle$    & $\langle$3, 0$\rangle$                      \\
                                                                                                                                                 & mathgen                 & 735                           & $\langle$100, 100$\rangle$        & 5                                                                                                                             & $\langle$5, 5$\rangle$                  & $\langle$4, 5$\rangle$                          & $\langle$1, 0$\rangle$                      & 5                                       & 5                                               & 0                                           & 11                                      & 11                                              & 0                                           & 96.4            & 98.6             & 82                      & $\langle$79, 82$\rangle$    & $\langle$3, 0$\rangle$                      \\
                                                                                                                                                 & rbt                         & 366                           & $\langle$100, 100$\rangle$        & 10                                                                                                                            & $\langle$10, 10$\rangle$                & $\langle$10, 10$\rangle$                        & $\langle$0, 0$\rangle$                      & 10                                      & 10                                              & 0                                           & 5                                       & 5                                               & 0                                           & 87.1            & 88.4             & 27                      & $\langle$27, 27$\rangle$    & $\langle$0, 0$\rangle$                      \\
                                                                                                                                                 & strsim                   & 654                           & $\langle$100, 100$\rangle$        & 19                                                                                                                            & $\langle$19, 19$\rangle$                & $\langle$19, 19$\rangle$                        & $\langle$0, 0$\rangle$                      & 19                                      & 19                                              & 0                                           & 64                                      & 64                                              & 0                                           & 88.8            & 94.1             & 50                      & $\langle$50, 50$\rangle$    & $\langle$0, 0$\rangle$                      \\
                                                                                                                                                 & toml                       & 1206                          & $\langle$100, 100$\rangle$        & 12                                                                                                                            & $\langle$12, 12$\rangle$                & $\langle$10, 12$\rangle$                        & $\langle$2, 0$\rangle$                      & 12                                      & 12                                              & 0                                           & 150                                     & 150                                             & 0                                           & 72.6            & 83.2             & 47                      & $\langle$43, 47$\rangle$    & $\langle$4, 0$\rangle$                      \\ 
\hline
\rowcolor[rgb]{0.8,0.902,0.902} \textbf{Total}                                                                                                   &                                        & 4077                          & $\langle$100, 100$\rangle$        & 74                                                                                                                            & $\langle$74, 74$\rangle$                & $\langle$69, 74$\rangle$                        & $\langle$5, 0$\rangle$                      & 74                                      & 74                                              & 0                                           & 306                                     & 306                                             & 0                                           & 86.3            & 91.5             & 302                     & $\langle$291, 302$\rangle$  & $\langle$11, 0$\rangle$                     \\ 
\hline
\multirow{4}{*}{\begin{tabular}[c]{@{}c@{}}SWE-agent\\(C$\rightarrow$Rust)\end{tabular}}                                 & \crust-$\alpha$ & 22961                         & $\langle$40, 40$\rangle$          & 166                                                                                                                           & $\langle$153, 166$\rangle$              & $\langle$130, 146$\rangle$                      & $\langle$23, 20$\rangle$                    & -                                       & -                                               & -                                           & 493                                     & 493                                             & 0                                           & 68.2            & 75.7             & 673                     & -                           & -                                           \\
                                                                                                                                                 & \crust-$\beta$  & 66704                         & $\langle$0, 49$\rangle$           & 321                                                                                                                           & $\langle$0, 320$\rangle$                & $\langle$0, 295$\rangle$                        & $\langle$0, 25$\rangle$                     & -                                       & -                                               & -                                           & 1118                                    & 1114                                            & 4                                           & 57.9            & 75.2             & 1900                    & -                           & -                                           \\
                                                                                                                                                 & \crust-$\sigma$ & 3894                          & $\langle$1, 0$\rangle$            & 1                                                                                                                             & $\langle$1, 0$\rangle$                  & $\langle$1, 0$\rangle$                          & $\langle$0, 0$\rangle$                      & -                                       & -                                               & -                                           & 53                                      & 51                                              & 2                                           & 4.3             & 67.4             & 41                      & -                           & -                                           \\
                                                                                                                                                 & \crust-$\gamma$ & 15169                         & $\langle$0, 0$\rangle$            & 135                                                                                                                           & $\langle$0, 0$\rangle$                  & $\langle$0, 0$\rangle$                          & $\langle$0, 0$\rangle$                      & -                                       & -                                               & -                                           & 274                                     & 270                                             & 4                                           & 27.0            & 44.7             & 572                     & -                           & -                                           \\ 
\hline
\rowcolor[rgb]{0.8,0.902,0.902} \textbf{\textbf{Total}}                                                                                          &                                        & 108728                        & $\langle$41, 89$\rangle$          & 623                                                                                                                           & $\langle$154, 486$\rangle$              & $\langle$131, 441$\rangle$                      & $\langle$23, 45$\rangle$                    & -                                       & -                                               & -                                           & 1938                                    & 1928                                            & 10                                          & 39.3            & 65.7             & 3186                    & -                           & -                                           \\ 
\hline\hline
\rowcolor[rgb]{0.902,0.902,0.902} \textbf{Total}                                                                                                 &                                        & 233057                        & $\langle$96.9, 99.4$\rangle$      & 2107                                                                                                                          & $\langle$1068, 1970$\rangle$            & $\langle$542, 1822$\rangle$                     & $\langle$526, 148$\rangle$                  & 1484                                    & 1456                                            & 28                                          & 3642                                    & 3629                                            & 13                                          & 70.8            & 85.4             & 4583                    & $\langle$935, 1366$\rangle$ & $\langle$462, 31$\rangle$                   \\
\hline
\end{tabular}
    }
    \label{table:effectiveness}
    \vspace{-10pt}
\end{table*}

\subsubsection{Implementation} For validating translations, \approach uses Rust $1.92.0$, Python $3.12.9$, Java $21.0.7$, Node $22.16.0$, GCC $12.2.0$, and Go $1.24.4$. We use Anthropic's Claude Code $2.1.19$~\cite{claudecode} \revision{as our agent SDK} for the agentic workflow discussed in \S\ref{sec:approach}. \revision{Specifically, \approach uses Claude Code to run the main agent loop for each of the four agents discussed in \S\ref{sec:approach}. Moreover, we also use Claude Code's basic tools (e.g., \texttt{\small Read}, \texttt{\small Write}, etc.), however, \approach implements the MCP tools discussed in \S\ref{subsec:tools}.}

% \subsubsection{Implementation} For validating translations, \approach uses Rust $1.92.0$~\cite{rustlang}, Python $3.12.9$~\cite{pythonlang}, Java $21.0.7$~\cite{javalang}, Node $22.16.0$~\cite{nodejs}, GCC $12.2.0$~\cite{gcc}, and Go $1.24.4$~\cite{golang}. We use Anthropic's Claude Code $2.1.19$~\cite{claudecode} for the agentic workflow discussed in \S\ref{sec:approach}.

\subsection{RQ1: Effectiveness of \approach}
\label{subsec:rq1}

Table~\ref{table:effectiveness} shows the results of \approach and other techniques in repository-level code translation and validation. We assess effectiveness from three different aspects: (1) \emph{Syntactic Correctness} (\S\ref{subsubsec:rq1:syntactic-correctness}), (2) \emph{Test Validation} (\S\ref{subsubsec:rq1:test-validation}), and (3) \emph{Function Validation} (\S\ref{subsubsec:rq1:functional-validation}).

\vspace{-5pt}
\subsubsection{Syntactic Correctness}
\label{subsubsec:rq1:syntactic-correctness}

\approach achieves an overall Compilation Success (CS) of $99.4\%$ across all projects, surpassing competing techniques which attain $96.9\%$. For projects in \oxidizer, \alphatrans, and \skel, both \approach and existing techniques produce $100\%$ compilable code. The most significant improvement is observed in \crust, where \approach generates compilable translations for $\frac{89}{100}$ projects--an improvement of $48$ projects over SWE-agent. This improvement is particularly notable given the difficulty of C$\rightarrow$Rust translation, especially with respect to memory management and ownership semantics. For instance, the following function \texttt{\small writechar} from \texttt{\small printf} is translated properly by SWE-agent; however, its call sites inconsistently use \texttt{\small int} and \texttt{\small char} as the first argument. While this behavior is acceptable in C, where a \texttt{\small char} is represented as an \texttt{\small int} in memory, it is invalid in Rust, where these types are distinct and thus lead to compilation errors. In contrast, \approach consistently invokes \texttt{\small writechar} with the appropriate argument type \texttt{\small char}.

% \vspace{5pt}
\noindent
\begin{minipage}{0.49\linewidth}
\begin{minted}[frame=lines,framesep=1mm,baselinestretch=0.5, fontsize=\scriptsize, breaklines, breakanywhere, linenos,numbersep=2pt]{c}
------------ C SOURCE CODE ------------
int	writechar(char c, int *len) {
    return ((*len)++, write(1, &c, 1));
}
\end{minted}
\end{minipage}\hfill
\begin{minipage}{.49\linewidth}
\begin{minted}[escapeinside=||, frame=lines,framesep=1mm,baselinestretch=0.5, fontsize=\scriptsize, breaklines, breakanywhere, linenos,numbersep=2pt]{rust}
----------- RUST TRANSLATION ----------
pub fn writechar(c: char, len: &mut i32) -> i32 {
    *len += 1; ...
}
\end{minted}
\end{minipage}
\vspace{-5pt}

\subsubsection{Test Validation}
\label{subsubsec:rq1:test-validation}

To evaluate the functional equivalence of translations, we execute source PL developer tests and measure the number of tests executed and passing. If existing tests do not cover certain functions, \approach generates tests to validate them.

\textit{\textbf{Validated Developer Tests.}} To fairly evaluate translations across all projects and to eliminate the threat of incorrectly translated tests, we use the validated test suites provided in the artifacts of prior tools. \revision{We only used validated test suites after the execution of \approach to ensure they remain hidden during experiments}. Because \oxidizer does not translate tests, we manually translated and validated the Go tests into Rust. Multi-column \emph{Validated Developer Tests} in Table~\ref{table:effectiveness} shows the number of executed, passing, and failing tests for existing tools and \approach. As corroborated in the table, \approach substantially improves test pass rate (TPR), passing $\frac{1{,}822}{2{,}107}$ tests ($86.5\%$), compared to only $\frac{542}{2{,}107}$ tests ($25.7\%$) for competing techniques, improving TPR by $60.8\%$. In particular, \approach achieves $100\%$ TPR compared to \oxidizer and \skel which achieve $67.2\%$ and $93.2\%$, respectively. For the \crust benchmark, our comparison is restricted to the $40$ projects for which both \approach and SWE-agent produce compilable translations (\crust-$\alpha$). Out of $166$ available tests, \approach executes and passes $146$ tests ($88.0\%$ TPR), while SWE-agent achieves a TPR of $78.3\%$. The largest gain is observed in \alphatrans, where \approach passes $1{,}078$ tests compared to only $188$ tests by \alphatrans's compositional approach, an improvement of $75.4\%$. The reduced performance of \alphatrans is primarily due to its limited number of executed tests caused by test collection errors; for example, in \texttt{\small cli} only $\frac{66}{381}$ tests are executed. The following snippet illustrates one such problematic translation that is required by most test classes and leads to test collection errors. The Java code uses a \texttt{\small protected} constructor to restrict who can create \texttt{\small CommandLine} instances while still allowing controlled construction within the package or subclasses. By contrast, the Python translation replaces this with a runtime check that always raises a \texttt{\small TypeError} when \texttt{\small CommandLine} is instantiated directly, effectively making the class non-instantiable and changing the original design intent.

% \vspace{5pt}
\noindent
\begin{minipage}{0.49\linewidth}
\begin{minted}[frame=lines,framesep=1mm,baselinestretch=0.5, fontsize=\scriptsize, breaklines, breakanywhere, linenos,numbersep=2pt]{java}
----------- JAVA SOURCE CODE ----------
public class CommandLine implements Serializable {
    protected CommandLine() {}
}
\end{minted}
\end{minipage}\hfill
\begin{minipage}{.49\linewidth}
\begin{minted}[escapeinside=||, frame=lines,framesep=1mm,baselinestretch=0.5, fontsize=\scriptsize, breaklines, breakanywhere, linenos,numbersep=2pt]{py}
---------- PYTHON TRANSLATION ---------
class CommandLine:
    def __init__(self) -> None:
        if type(self) is CommandLine:
            raise TypeError("Error")
\end{minted}
\end{minipage}
% \vspace{5pt}

\textit{\textbf{\approach Translated Developer Tests.}} In addition to evaluating translations using validated developer tests, we also execute developer tests translated by \approach to assess its capability in test translation. A detailed analysis of translated test quality is provided in \S\ref{subsec:rq2}. Multi-column \emph{\approach Translated Developer Tests} in Table~\ref{table:effectiveness} summarize these results. Across $1{,}484$ translated tests, excluding \crust where test translation is not required \revision{as their authors provide validated tests in Rust and include it in their prompts}, \approach executes and passes $1{,}456$ tests ($98.1\%$), with only $28$ failures. Except for \alphatrans, \approach can correctly translate and produce tests equivalent to those in the source PL in \oxidizer and \skel mostly because they have simpler test logic. We further analyzed the discrepancies in test failures between validated developer tests and translated ones. Specifically, we identified incorrectly validated tests by the authors of \alphatrans as shown below. This example is from \texttt{\small csv} project with $57$ test failures from validated developer tests, but none from \approach translated tests. The \texttt{\small printRecord1} invocation in \texttt{\small testJiraCsv249} takes two string arguments in source Java tests, but was incorrectly translated to take a list in Python and therefore fails. This test translated by \approach has the same semantics as Java and passes correctly, demonstrating its ability in automated test translation.

% \vspace{5pt}
\noindent
\begin{minipage}{0.49\linewidth}
\begin{minted}[frame=lines,framesep=1mm,baselinestretch=0.5, fontsize=\scriptsize, breaklines, breakanywhere, linenos,numbersep=2pt]{java}
------------ JAVA TEST CODE -----------
public void testJiraCsv249() {
    ...
    printer.printRecord1("foo \\", "bar");
    ...
}
\end{minted}
\end{minipage}\hfill
\begin{minipage}{.49\linewidth}
\begin{minted}[escapeinside=||, frame=lines,framesep=1mm,baselinestretch=0.5, fontsize=\scriptsize, breaklines, breakanywhere, linenos,numbersep=2pt]{py}
-------- ALPHATRANS TRANSLATION -------
def testJiraCsv249(self) -> None:
    ...
    printer.printRecord1(["foo \\", "bar"])

    ...
\end{minted}
\end{minipage}
% \vspace{5pt}

\begin{table*}[t]
    % \setlength{\tabcolsep}{1.7pt}
    % \scriptsize
    \centering
    \caption{Comparison between \approach translated tests and original source PL tests. Tuple entries indicate $\langle$Source Test, Translated Test$\rangle$. LoC: Lines of Code.}
    \vspace{-10pt}
    \resizebox{\linewidth}{!}{
        \begin{tabular}{c|c|c|c|c|c|c|c|c|c|c|c|c} 
\hline
\multirow{2}{*}{\textbf{Tool}}                                                                  & \multirow{2}{*}{\textbf{Project}} & \multirow{2}{*}{\begin{tabular}[c]{@{}c@{}}\textbf{\# }\\\textbf{Tests}\end{tabular}} & \multirow{2}{*}{\begin{tabular}[c]{@{}c@{}}\textbf{\# Tests}\\\textbf{Translated /}\\\textbf{Not Translated}\end{tabular}} & \multirow{2}{*}{\begin{tabular}[c]{@{}c@{}}\textbf{\# Tests w/ Matching}\\\textbf{/ Non-Matching}\\\textbf{\# Assertions}\end{tabular}} & \multirow{2}{*}{\begin{tabular}[c]{@{}c@{}}\textbf{\textbf{\# Total / Matching}}\\\textbf{\textbf{assertEqual Output}}\end{tabular}} & \multicolumn{4}{c|}{\textbf{Assertion Type Match (\%)}}                                                                                                                                                                                     & \multirow{2}{*}{\begin{tabular}[c]{@{}c@{}}\textbf{\textbf{\textbf{\textbf{Avg. Cosine}}\textbf{\textbf{}}}}\\\textbf{\textbf{\textbf{\textbf{Similarity}}}}\end{tabular}} & \multirow{2}{*}{\textbf{Avg. LoC}} & \multirow{2}{*}{\begin{tabular}[c]{@{}c@{}}\textbf{Avg. \# Method}\\\textbf{Invocations}\end{tabular}}  \\ 
\cline{7-10}
                                                                                                &                                   &                                                                                       &                                                                                                                            &                                                                                                                                         &                                                                                                                                      & \begin{tabular}[c]{@{}c@{}}\textbf{Assert}\\\textbf{Equal}\end{tabular} & \begin{tabular}[c]{@{}c@{}}\textbf{Assert}\\\textbf{True}\end{tabular} & \begin{tabular}[c]{@{}c@{}}\textbf{Assert}\\\textbf{False}\end{tabular} & \textbf{Other} &                                                                                                                                                                            &                                    &                                                                                                         \\ 
\hline
\multirow{6}{*}{\begin{tabular}[c]{@{}c@{}}\oxidizer\\(Go$\rightarrow$Rust)\end{tabular}}       & checkdigit                        & 36                                                                                    & 36/0                                                                                                                       & 36/0                                                                                                                                    & 45/45                                                                                                                                & 100                                                                     & -                                                                      & -                                                                       & -              & 0.94                                                                                                                                                                       & $\langle$24.42, 24.58$\rangle$     & $\langle$7.14, 5.33$\rangle$                                                                            \\
                                                                                                & go-edlib                          & 36                                                                                    & 36/0                                                                                                                       & 36/0                                                                                                                                    & 45/45                                                                                                                                & 84.44                                                                   & -                                                                      & -                                                                       & -              & 0.92                                                                                                                                                                       & $\langle$30.78, 51$\rangle$        & $\langle$7.61, 9.50$\rangle$                                                                            \\
                                                                                                & histogram                         & 2                                                                                     & 2/0                                                                                                                        & 2/0                                                                                                                                     & 11/11                                                                                                                                & 100                                                                     & -                                                                      & -                                                                       & -              & 0.96                                                                                                                                                                       & $\langle$23.50, 16.50$\rangle$     & $\langle$24, 16.50$\rangle$                                                                             \\
                                                                                                & nameparts                         & 26                                                                                    & 26/0                                                                                                                       & 26/0                                                                                                                                    & 51/51                                                                                                                                & 100                                                                     & -                                                                      & -                                                                       & -              & 0.94                                                                                                                                                                       & $\langle$11.96, 6.31$\rangle$      & $\langle$6.15, 3.81$\rangle$                                                                            \\
                                                                                                & stats                             & 121                                                                                   & 121/0                                                                                                                      & 121/0                                                                                                                                   & 150/150                                                                                                                              & 91.15                                                                   & -                                                                      & -                                                                       & -              & 0.85                                                                                                                                                                       & $\langle$16.82, 9.44$\rangle$      & $\langle$8.12, 6$\rangle$                                                                               \\
                                                                                                & textrank                          & 8                                                                                     & 8/0                                                                                                                        & 8/0                                                                                                                                     & 12/12                                                                                                                                & 100                                                                     & -                                                                      & -                                                                       & -              & 0.91                                                                                                                                                                       & $\langle$13.75, 15.25$\rangle$     & $\langle$10.88, 11.88$\rangle$                                                                          \\ 
\hline
\rowcolor[rgb]{0.8,0.902,0.902} \textbf{Total}                                                  &                                   & 229                                                                                   & 229/0                                                                                                                      & 229/0                                                                                                                                   & 314/314                                                                                                                              & 95.93                                                                   & -                                                                      & -                                                                       & -              & 0.92                                                                                                                                                                       & $\langle$20.21, 20.51$\rangle$     & $\langle$10.65, 8.84$\rangle$                                                                           \\ 
\hline
\multirow{4}{*}{\begin{tabular}[c]{@{}c@{}}\alphatrans\\(Java$\rightarrow$Python)\end{tabular}} & cli                               & 381                                                                                   & 381/0                                                                                                                      & 381/0                                                                                                                                   & 452/452                                                                                                                              & 99.61                                                                   & 99.68                                                                  & 98.37                                                                   & 97.87          & 0.90                                                                                                                                                                       & $\langle$12.41, 11.59$\rangle$     & $\langle$13.30, 12.63$\rangle$                                                                          \\
                                                                                                & csv                               & 298                                                                                   & 298/0                                                                                                                      & 292/6                                                                                                                                   & 207/207                                                                                                                              & 100                                                                     & 81.82                                                                  & 84.31                                                                   & 92.86          & 0.90                                                                                                                                                                       & $\langle$11.84, 8.94$\rangle$      & $\langle$12.20, 10.46$\rangle$                                                                          \\
                                                                                                & fileupload                        & 39                                                                                    & 39/0                                                                                                                       & 39/0                                                                                                                                    & 37/37                                                                                                                                & 100                                                                     & 100                                                                    & 80                                                                      & 100            & 0.87                                                                                                                                                                       & $\langle$6.74, 7.38$\rangle$       & $\langle$5.87, 6.36$\rangle$                                                                            \\
                                                                                                & validator                         & 463                                                                                   & 463/0                                                                                                                      & 458/5                                                                                                                                   & 374/374                                                                                                                              & 98.52                                                                   & 99.28                                                                  & 99.49                                                                   & 90.72          & 0.89                                                                                                                                                                       & $\langle$17.69, 17.23$\rangle$     & $\langle$18.78, 18.30$\rangle$                                                                          \\ 
\hline
\rowcolor[rgb]{0.8,0.902,0.902} \textbf{Total}                                                  &                                   & 1181                                                                                  & 1181/0                                                                                                                     & 1170/11                                                                                                                                 & 1070/1070                                                                                                                            & 99.53                                                                   & 95.20                                                                  & 90.54                                                                   & 95.36          & 0.89                                                                                                                                                                       & $\langle$12.17, 11.29$\rangle$     & $\langle$12.54, 11.94$\rangle$                                                                          \\ 
\hline
\multirow{8}{*}{\begin{tabular}[c]{@{}c@{}}\skel\\(Python$\rightarrow$JavaScript)\end{tabular}} & bst                               & 11                                                                                    & 11/0                                                                                                                       & 11/0                                                                                                                                    & 74/74                                                                                                                                & 100                                                                     & -                                                                      & -                                                                       & -              & 0.91                                                                                                                                                                       & $\langle$22.27, 22.64$\rangle$     & $\langle$5.73, 12$\rangle$                                                                              \\
                                                                                                & colorsys                          & 2                                                                                     & 2/0                                                                                                                        & 2/0                                                                                                                                     & 39/39                                                                                                                                & 100                                                                     & -                                                                      & -                                                                       & -              & 0.93                                                                                                                                                                       & $\langle$67, 65$\rangle$           & $\langle$45, 44$\rangle$                                                                                \\
                                                                                                & heapq                             & 8                                                                                     & 8/0                                                                                                                        & 8/0                                                                                                                                     & 29/29                                                                                                                                & 100                                                                     & -                                                                      & -                                                                       & -              & 0.90                                                                                                                                                                       & $\langle$13.12, 13.38$\rangle$     & $\langle$11.25, 10.25$\rangle$                                                                          \\
                                                                                                & html                              & 7                                                                                     & 7/0                                                                                                                        & 7/0                                                                                                                                     & 19/19                                                                                                                                & 100                                                                     & -                                                                      & -                                                                       & -              & 0.92                                                                                                                                                                       & $\langle$24.71, 22.71$\rangle$     & $\langle$20.29, 19.29$\rangle$                                                                          \\
                                                                                                & mathgen                           & 5                                                                                     & 5/0                                                                                                                        & 5/0                                                                                                                                     & 163/163                                                                                                                              & 100                                                                     & -                                                                      & -                                                                       & -              & 0.93                                                                                                                                                                       & $\langle$47, 51$\rangle$           & $\langle$49, 42$\rangle$                                                                                \\
                                                                                                & rbt                               & 10                                                                                    & 10/0                                                                                                                       & 10/0                                                                                                                                    & 16/16                                                                                                                                & 100                                                                     & -                                                                      & -                                                                       & -              & 0.91                                                                                                                                                                       & $\langle$17.90, 19.60$\rangle$     & $\langle$20.90, 17.10$\rangle$                                                                          \\
                                                                                                & strsim                            & 19                                                                                    & 19/0                                                                                                                       & 19/0                                                                                                                                    & 150/150                                                                                                                              & 100                                                                     & -                                                                      & -                                                                       & -              & 0.93                                                                                                                                                                       & $\langle$18.74, 17.21$\rangle$     & $\langle$22.84, 21.84$\rangle$                                                                          \\
                                                                                                & toml                              & 12                                                                                    & 12/0                                                                                                                       & 12/0                                                                                                                                    & 17/17                                                                                                                                & 100                                                                     & -                                                                      & -                                                                       & -              & 0.90                                                                                                                                                                       & $\langle$8.67, 8.75$\rangle$       & $\langle$6.08, 7$\rangle$                                                                               \\ 
\hline
\rowcolor[rgb]{0.8,0.902,0.902} \textbf{Total}                                                  &                                   & 74                                                                                    & 74/0                                                                                                                       & 74/0                                                                                                                                    & 507/507                                                                                                                              & 100                                                                     & -                                                                      & -                                                                       & -              & 0.92                                                                                                                                                                       & $\langle$27.43, 27.54$\rangle$     & $\langle$22.64, 21.69$\rangle$                                                                          \\ 
\hline\hline
\rowcolor[rgb]{0.902,0.902,0.902} \textbf{Total}                                                &                                   & 1484                                                                                  & 1484/0                                                                                                                     & 1473/11                                                                                                                                 & 1891/1891                                                                                                                            & 98.54                                                                   & 95.20                                                                  & 90.54                                                                   & 95.36          & 0.91                                                                                                                                                                       & $\langle$21.63, 21.58$\rangle$     & $\langle$16.40, 15.24$\rangle$                                                                          \\
\hline
\end{tabular}
    }
    \label{table:test-comparison}
    \vspace{-10pt}
\end{table*}

\textit{\textbf{\approach Generated Tests.}} A major limitation of source PL developer tests is their low coverage and the inability to validate unexercised translations. For instance, the \texttt{\small fileupload} project in \alphatrans has a line coverage of only $38.7\%$, leaving the remaining $61.3\%$ of translated lines unvalidated. To mitigate this, \approach generates additional tests for each project (\S\ref{subsec:validator-agent}). This capability addresses a fundamental limitation in existing validation approaches: the quality of validation is inherently bounded by the coverage of the original test suite. Projects with low test coverage may have large portions of translated code that remain unvalidated, potentially hiding translation bugs. Multi-column \emph{\approach Generated Tests} indicates generated tests executed, passing, and failing. Across all benchmarks, \approach generates $3{,}642$ tests and achieves a TPR of $99.6\%$ ($3{,}629$ passing, $13$ failing). The high pass rate on generated tests indicates that \approach's test generation component produces valid tests that correctly exercise the translated code. Moreover, the generated tests increase test coverage significantly: on average, test coverage improves from $70.8\%$ to $85.4\%$, representing a $14.6\%$ improvement. The coverage improvement is particularly valuable for the \alphatrans benchmark, where the original projects have varying levels of test coverage. By generating additional tests, \approach validates code paths that were previously unvalidated, increasing confidence in the correctness of the translated implementation.

\revision{Moreover, we manually investigated the quality and effectiveness of generated tests by \approach. To perform the study, we created a benchmark of $502$ tests uniformly sampled from all four existing techniques. Two human investigators independently analyzed each generated test and the method \approach used to determine coverage gap with an inter-rater agreement of $100\%$. On average, our manual investigation show that \approach can correctly generate an effective test which exercises previously uncovered code $71.1\%$ of the time. To determine coverage gap, our analysis of agent trajectories indicate that for larger projects like \alphatrans, \approach generates internal scripts to extract uncovered code. For smaller projects in other benchmarks, it simply relies on reading source and test code. The details of our human study with comments from investigators is publicly available~\cite{website}.}

\vspace{-10pt}
\subsubsection{Function Validation}
\label{subsubsec:rq1:functional-validation}

So far, we only used test validation for evaluating the functional correctness of translations. However, the problem of \emph{test translation coupling effect} discussed in \alphatrans~\cite{ibrahimzada2025alphatrans} still exists. That is, a translation issue in one method casts a shadow in validating the translation of the other methods. Consequently, test validation alone is not a good metric for evaluating functional correctness, as it can heavily favor one technique over the others. To address this, we evaluate each translated function independently as an alternative way to measure correctness.

For benchmarks where function-level validation is possible, we evaluate whether each translated function produces correct output when invoked with the same inputs as the original function. Since \crust only performs test validation, we excluded it from function validation. Multi-column \emph{Function Validation} shows the results of this evaluation. Across $1{,}397$ functions, \approach achieves successful validation for $1{,}366$ ($97.8\%$) compared to $935$ ($66.9\%$) for competing techniques. This improvement of $30.9\%$ demonstrates that test validation ($60.8\%$ improvement in TPR) alone can be unreliable and produce inflated improvements. For instance, let's consider the following example from \texttt{\small nameparts} project in \oxidizer which validates the \texttt{\small Parse} function. When doing test validation (left), the test only checks if the function panics on the input \texttt{\small "I am a Popsicle"}. However, when performing function validation (right), the test goes beyond checking for panic, and asserts the parsed properties, i.e., \texttt{\small FullName}. As a result, test validation validates the \texttt{\small Parse} function as correct since it does not panic, however, function validation fails because \texttt{\small FullName} is not parsed correctly, demonstrating function validation's more rigorous evaluation.

% \vspace{5pt}
\noindent
\begin{minipage}{0.49\linewidth}
\begin{minted}[frame=lines,framesep=1mm,baselinestretch=0.5, fontsize=\scriptsize, breaklines, breakanywhere, linenos,numbersep=2pt]{rust}
----------- TEST VALIDATION -----------
fn TestObviouslyBadName() {
    let result = std::panic::catch_unwind(|| {
        Parse("I am a Popsicle".to_string())
    });
    assert!(result.is_ok(), "Parse should not panic on invalid input");
}
\end{minted}
\end{minipage}\hfill
\begin{minipage}{.49\linewidth}
\begin{minted}[escapeinside=||, frame=lines,framesep=1mm,baselinestretch=0.5, fontsize=\scriptsize, breaklines, breakanywhere, linenos,numbersep=2pt]{rust}
--------- FUNCTION VALIDATION ---------
pub fn parse__unit_test() {
    ...    
    let result = Parse(input_name);
    assert_eq!(result.FullName, expected_result["FullName"].as_str().unwrap_or(""), 
        "Test case {}: FullName mismatch", i);
    ...
}
\end{minted}
\end{minipage}
% \vspace{5pt}

\subsection{RQ2: Test Translation}
\label{subsec:rq2}

Table~\ref{table:test-comparison} shows the results of \approach's test translation capabilities, comparing translated tests against their original source PL counterparts. This evaluation covers $1{,}484$ tests across three PL pairs, excluding the \crust benchmark, as it does not require test translation. \approach successfully translates all $1{,}484$ source tests to their target PLs, achieving $100\%$ translation rate across all benchmarks, demonstrating the ability of \approach's Validator Agent to ensure all tests are properly translated with no empty test logic. For translated tests to be \revision{structurally similar}, they should preserve the same number of assertions as the original tests. Across all benchmarks, $\frac{1{,}473}{1{,}484}$ tests ($99.3\%$) have matching assertion counts between source and translated PLs. Only $11$ tests exhibit non-matching counts, all in \alphatrans projects. These cases occur when source tests contain a significantly large number of assert statements (e.g., $50+$) due to hallucination.

\begin{figure*}[t]
    \centering
    \includegraphics[width=\linewidth]{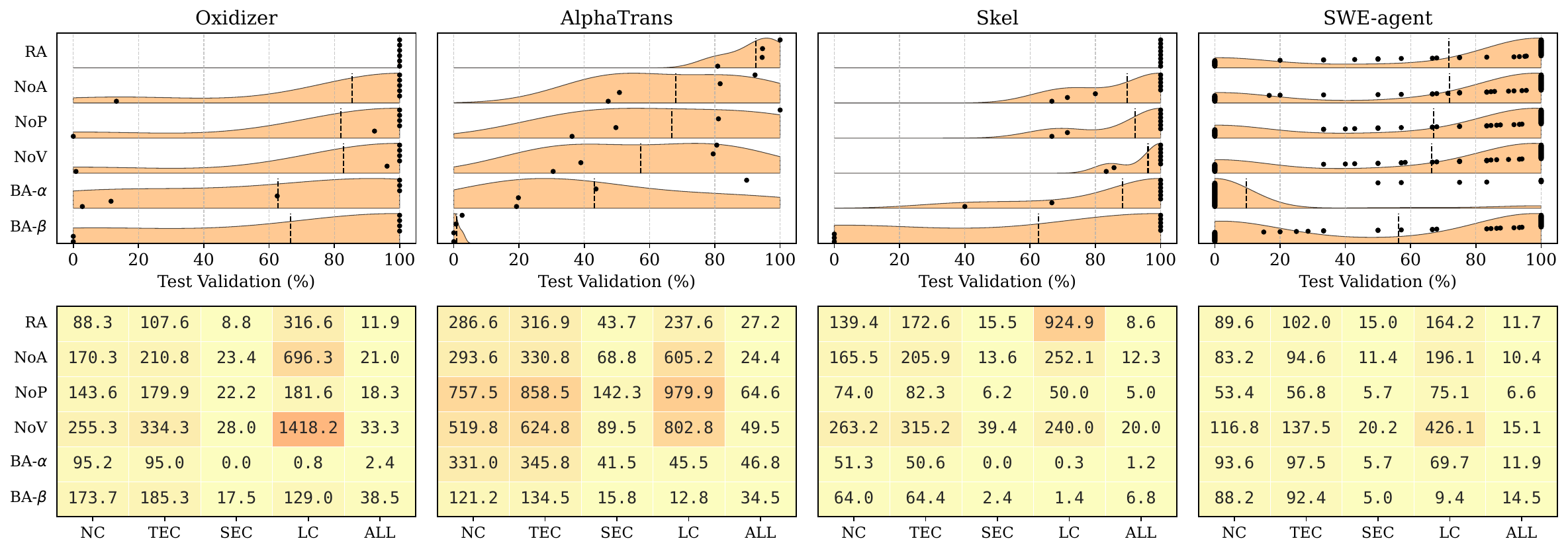}
    %\vspace{-10pt}
    \caption{Impact of different agents on translation effectiveness and agent trajectories. RA: \approach, NoA: No Analyzer, NoP: No Planning, NoV: No Validator, BA-$\alpha$: Base Agent with Prompt Condensation, BA-$\beta$: Base Agent with Prompt Concatenation, NC: Node Count, TEC: Temporal Edge Count, SEC: Structural Edge Count, LC: Loop Count, ALL: Average Loop Length.}
    \label{fig:ablation}
    %\vspace{-10pt}
\end{figure*}

For \texttt{\small assertEqual}-style assertions, we evaluate whether translated tests have the same expected values as original tests. Specifically, we check if expected outputs are similar for four types, \texttt{\small string, int, float, bool}. \approach achieves $100\%$ matching on \texttt{\small assertEqual} outputs, with all $1{,}891$ assertions producing equivalent expected values in target PL. This metric is critical because \texttt{\small assertEqual} assertions directly validate functional correctness---if a translated test expects a different output value, it would fail even on a correct translation. We also evaluate whether \approach preserves semantic types of assertions during translation. For instance, we check if \texttt{\small assertEqual(a, b)} in Java is translated to \texttt{\small assertEqual(a, b)} or \texttt{\small assertTrue(a)} when \texttt{\small b} is a boolean in Python. Multi-column \emph{Assertion Type Match} shows the results of this evaluation. For \texttt{\small assertEqual} assertions, \approach achieves $98.54\%$ match rate across all benchmarks. Specifically, \oxidizer achieves $95.93\%$ and \skel achieves $100\%$. For \alphatrans, which tests a broader variety of assertion types due to JUnit's rich assertion library, \approach achieves: $99.53\%$ for \texttt{\small assertEqual}, $95.20\%$ for \texttt{\small assertTrue}, $90.54\%$ for \texttt{\small assertFalse}, and $95.36\%$ for other assertions including \texttt{\small assertNull} and \texttt{\small assertThrows}. The lower match rate for \texttt{\small assertFalse} ($90.54\%$) is because of the translation of certain Java assertion idioms to semantically equivalent but syntactically different Python expressions. For instance, \texttt{\small assertFalse(list.isEmpty())} in Java is translated to \texttt{\small assert len(list) > 0} in Python, which tests the same condition but uses a different assertion pattern.

Moreover, we evaluate semantic similarity between source and translated tests using cosine similarity computed over code embeddings generated by \texttt{\small Qwen/Qwen3-Embedding-0.6B}~\cite{qwen-embedding}. This metric captures the degree to which translated tests preserve structural and logical patterns, independent of surface-level syntactic differences between languages. Across all benchmarks, \approach achieves an average cosine similarity of $0.91$, indicating high semantic preservation. We also compare structural characteristics using lines of code (LoC) and method invocation counts. On average, source tests contain $21.63$ LoC while translated tests contain $21.58$ LoC, a difference of less than $1\%$. This alignment indicates that \approach produces translations of comparable complexity without code bloat or oversimplification. For method invocations, source tests average $16.40$ invocations while translated tests average $15.24$, a reduction of $7\%$. This reduction is attributed to idiomatic differences between testing frameworks.

\revision{Beside non-functional metrics discussed above, we also performed manual investigation to assess the quality of translated tests. To evaluate tests across multiple benchmarks, we sampled $100$ translated tests from \alphatrans and \oxidizer, and used all $74$ tests from \skel. To do the human study, we first leveraged two state-of-the-art LLMs, namely, \texttt{\small claude-sonnet-5} and \texttt{\small claude-opus-4-8} to perform initial analysis on the quality of the translated tests. Two human investigators then manually checked their analysis to ensure correctness and met with each other to resolve conflicts. The results from our human study of translated tests indicate \approach achieves $88\%$ accuracy with an inter-rater agreement of $96.7\%$. Please refer to our artifacts for more comprehensive comments by each manual investigator~\cite{website}.}

\subsection{RQ3: Ablation Study}
\label{subsec:rq3}

Figure~\ref{fig:ablation} presents the results of our ablation studies, evaluating the contribution of each component in \approach. We compare \approach (RA) against five ablated configurations: No Analyzer (NoA), No Planning (NoP), No Validator (NoV), Base Agent with Prompt Condensation (BA-$\alpha$), and Base Agent with Prompt Concatenation (BA-$\beta$). The top portion of Figure~\ref{fig:ablation} shows test validation percentages and their distribution across all four benchmarks, while the bottom portion presents trajectory analysis using process-centric metrics, demonstrating agent behavior patterns.

\subsubsection{Impact of Individual Agents}

Removing the analyzer agent (NoA) results in decreased test validation performance ($\downarrow$$22.7\%$) across all benchmarks. Without foundational analysis of the source project structure and third-party library dependencies, the translator and validator agents must repeatedly explore the codebase, exhausting the context window and leading to inefficient interactions. The removal of the planning agent (NoP) demonstrates even more significant performance degradation ($\downarrow$$25.3\%$), as the translator agent lacks structured guidance for translation ordering, often resulting in repeated attempts. Finally, removing the validator agent (NoV) leads to the highest decrease in test validation ($\downarrow$$30.3\%$), as translation errors accumulate without dedicated test execution and diagnostic feedback.

\vspace{-5pt}
\subsubsection{Comparison with Base Agents}

The base agent configurations represent \approach without specialized agents. BA-$\alpha$ condenses the entire translation task into a single compact prompt, while BA-$\beta$ concatenates all \approach prompts into a large prompt. Both perform significantly worse than \approach across all benchmarks, by as much as $61.2\%$ for BA-$\alpha$ and $62.4\%$ for BA-$\beta$. These results demonstrate that simply providing an LLM with all available information does not yield effective translation---the structured decomposition into analysis, planning, translation, and validation phases is essential.

\subsubsection{Trajectory Analysis}

To perform a process-centric analysis of agent trajectories, we use Graphectory~\cite{liu2026process} \revision{metrics}. \revision{These metrics carry the same interpretation in code translation similar to SWE-bench for fixing software engineering bugs, e.g., a higher node count (NC), temporal edge count (TEC), or loop count (LC) indicates the agent takes more actions to solve the problem}. Our objective is to show that test validation degradation alone is insufficient for evaluating ablations. The heatmaps in Figure~\ref{fig:ablation} reveal distinct patterns in agent behavior. \approach exhibits the most compact trajectories with the lowest average node and temporal edge counts, while achieving high test validation. The ablated configurations show progressively larger trajectory footprints as components are removed, up to $29\%$ and $27\%$ more NC and TEC. \revision{For example, NoV removes the validator agent and iterative feedback, depriving the translator agent of independent quality assurance. The NoV ablation shows higher NC, TEC, and LC because the translator agent performs both translation and validation within the same context window. This increased workload often leads to memory issues and higher LC. \approach avoids this by using the validator to simplify the translator’s task.} These process-centric metrics consistently show that \approach's multi-agent architecture provides structured guidance that prevents unnecessary exploration and repeated work.

% In summary, all three specialized agents contribute substantially to \approach's effectiveness, and removing any single component leads to significant degradation in both translation quality and efficiency.

\subsection{RQ4: Cost and Tool Usage Analysis}
\label{subsec:rq4}

\begin{figure}[t]
    \centering
    \includegraphics[width=\linewidth]{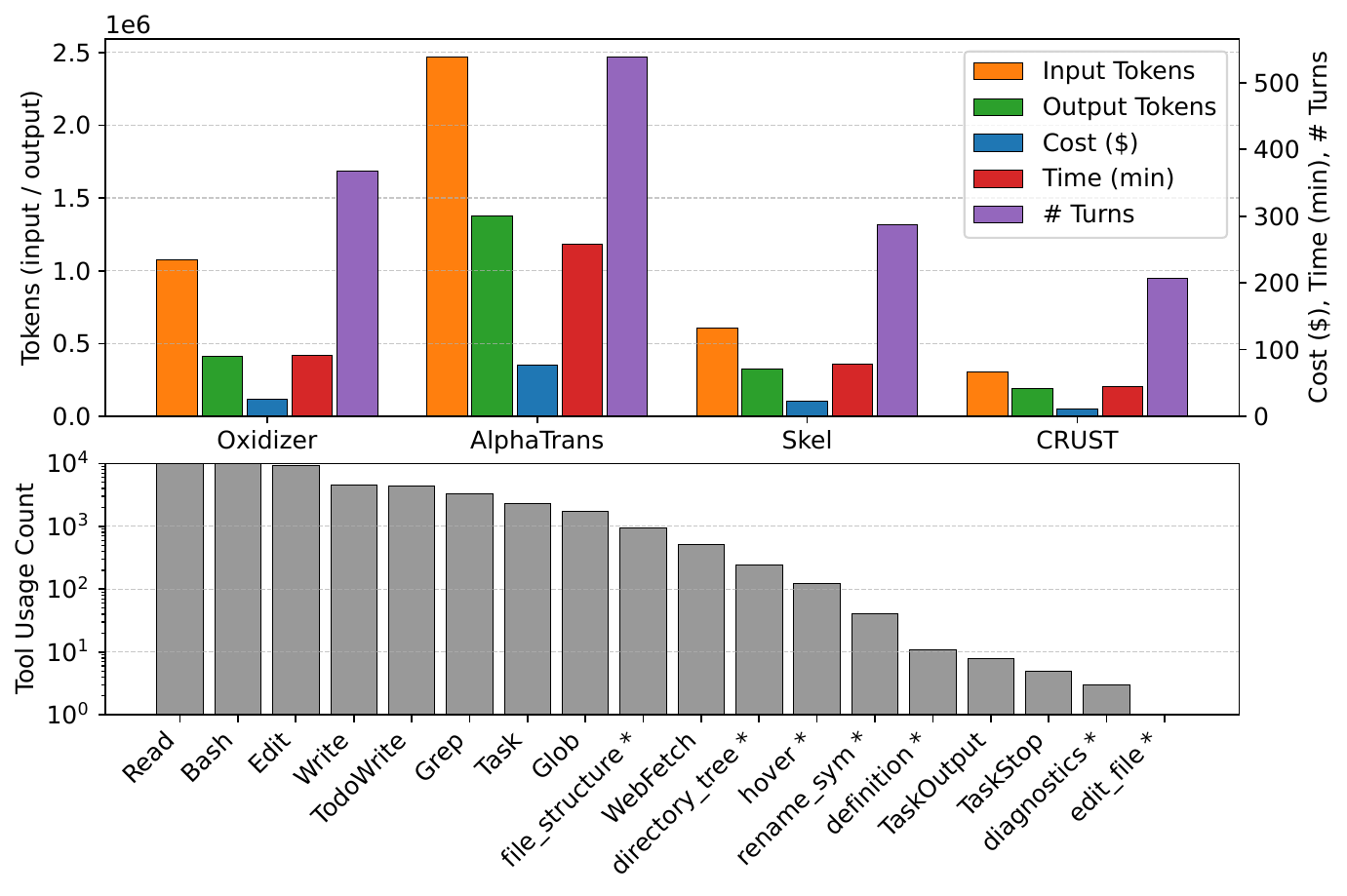}
    \vspace{-25pt}
    \caption{Cost and Tool Usage Analysis of \approach.}
    \label{fig:cost}
    \vspace{-15pt}
\end{figure}

Figure~\ref{fig:cost} presents \approach's computational costs per project and tool utilization patterns across all four benchmarks.

\subsubsection{Cost}

Project costs and token usage scale with complexity, ranging from \alphatrans ($2.5$M input/$1.4$M output tokens at $\$76$) and \oxidizer ($1.1$M input/$0.4$M output at $\$25$) to the more economical \skel ($0.6$M input/$0.3$M output at $\$20$) and \crust ($0.3$M input/$0.2$M output at $\$11$). Execution time scales linearly with project size: \alphatrans averages $258$ minutes per project, \oxidizer $92$ minutes, \skel $78$ minutes, and \crust $45$ minutes due to smaller individual project sizes. This predictable scaling enables users to estimate costs for new projects based on codebase size.

\revision{Moreover, Table~\ref{table:baseline_cost} shows the cost comparison between existing techniques and \approach averaged over number of translated projects. On average, it costs $\$15.3$ and $57$ minutes for \approach to translate and validate each project. The cost difference between \approach and existing techniques is justifiable as existing techniques can only translate \emph{a single PL pair} and have high development cost to generalize across other PLs.}

\begin{table}
    % \setlength{\tabcolsep}{2pt}
    % \scriptsize
    \centering
    \caption{\revision{Cost comparison of existing techniques and \approach averaged by number of projects translated.}}
    \vspace{-10pt}
    \resizebox{\columnwidth}{!}{
        \begin{tabular}{c|c|c|c|c} 
\hline
\revision{\textbf{Tool}}              & \revision{\textbf{Input Tokens}} & \revision{\textbf{Output Tokens}} & \revision{\textbf{Time (s)}} & \revision{\textbf{Cost (\$)}}  \\ 
\hline
\revision{\oxidizer}   & \revision{131085.0}              & \revision{44855.2}                & \revision{940.4}             & \revision{1.1}                 \\
\revision{\alphatrans} & \revision{2302094.0}             & \revision{296475.2}               & \revision{37503.4}           & \revision{11.4}                \\
\revision{\skel}       & \revision{45763.5}               & \revision{10733.6}                & \revision{450.8}             & \revision{0.3}                 \\
\revision{\crust}      & \revision{1060519.4}             & \revision{11899.7}                & \revision{455.1}             & \revision{4.0}                 \\
\revision{\approach}   & \revision{446829.7}              & \revision{257879.0}               & \revision{3433.2}            & \revision{15.3}                \\
\hline
\end{tabular}
    }
    \label{table:baseline_cost}
    \vspace{-10pt}
\end{table}

\vspace{-5pt}
\subsubsection{Tool Usage Distribution}

The bottom panel shows tool invocation patterns capped at $10K$. Core tools---\texttt{\small Read}, \texttt{\small Bash}, and \texttt{\small Edit}---are each invoked approximately $15{,}000$ times on average for examining code, executing commands, and modifying files. \texttt{\small Write} and \texttt{\small Grep} support file creation and search operations. \approach's LSP tools demonstrate targeted utilization: \texttt{\small file\_structure} (${\sim}1{,}000$ invocations), \texttt{\small hover} for type information (${\sim}150$), and semantic tools like \texttt{\small definition} (${\sim}40$) enable reliable code navigation.

% In summary, \approach is economically viable with costs scaling linearly with project complexity, positioning it as a practical alternative to heavily engineered neuro-symbolic approaches that require $3{,}843$--$19{,}052$ LoC of PL-specific implementation.

\section{Related Work}
\label{sec:related-work}

\noindent\textbf{Code Translation.} There are two main approaches for translating code from one PL to another: traditional rule-based transpiler techniques and LLMs. Transpiler tools like C2Rust~\cite{c2rust}, CxGo~\cite{c2go}, Sharpen~\cite{sharpen}, and Java2CSharp~\cite{java2csharp} translate code from C to Rust, C to Go, and Java to C\#, respectively. A series of statistical machine translation techniques~\cite{chen2018tree,nguyen2015divide,nguyen2013lexical,nguyen2014migrating} focus on translating Java to C\#. Deep learning approaches have also been applied for code translation~\cite{roziere2020unsupervised,roziere2022leveraging}. Recent advancements have focused on using LLMs for code translation~\cite{di2024codefuse,jiao2023evaluation,yin2024rectifier,yan2023codetransocean,tipirneni2024structcoder,pan2024lost}, which have demonstrated strong performance on synthetic benchmarks but limited effectiveness on real-world software projects. Furthermore, repository-level code translation has been studied for various PL pairs. \alphatrans~\cite{ibrahimzada2025alphatrans} \revision{is a neuro-symbolic technique with a deterministic pipeline---which includes program decomposition, type resolution, skeleton generation, etc. and is not autonomous like agentic approaches, e.g., \approach---to enable repository-level code translation and} translates Java to Python using open-source LLMs and GraalVM~\cite{graalvm} for isolated validation; \syzygy~\cite{shetty2024syzygy} targets C to Rust using GPT-4; \skel~\cite{wang2025skel} \revision{is also a neuro-symbolic technique that decomposes source project into smaller pieces, generates skeletons in target language and uses program execution traces for validation and} translates Python to JavaScript; \oxidizer~\cite{zhang2025oxidizer} \revision{is another neuro-symbolic approach that} employs type-driven techniques and language feature mapping to convert Go to Rust. \revision{\oxidizer extracts input/output pairs by executing source project tests and generates mock tests in target language for validating translations.} Some approaches have combined transpiler outputs with LLM-based translation~\cite{yang2025vert}, but their success is often limited by the availability and reliability of the underlying transpilers. \citet{nitin2024spectra} capture natural language specifications from source code to inform translation, while Yang et al.~\cite{yang2024exploring} utilize test cases to support the process. 

The rise of agent-based frameworks~\cite{chen2026unlocking,liu2026plan,chen2026can,liu2026large,xi2025the} has produced significant research and industrial interest in applying these architectures to a variety of software engineering challenges~\cite{jimenez2024swebench, yang2025swebench, chowdhury2024swebenchverified}. SWE-agent~\cite{yang2024sweagent} introduces a specialized agent-computer interface (ACI), enabling agents to interact with code repositories through file reading, editing, and execution of bash commands. \revision{SWE-agent from \crust~\cite{khatry2025crust} and MatchFixAgent~\cite{ibrahimzada2026matchfixagent} are the only autonomous agentic approach that has been adapted for code translation. Similar to \approach, they follows a cycle of reasoning and action to either only translate code, or only validate and repair the code translations. However, \approach automates the entire translation and validation pipeline into a multi-agent scaffold, improving the overall performance compared to these related approaches.}

% \vspace{3pt}

%\noindent\textbf{LLM Agents.} 
%\textsc{AutoCodeRover}~\cite{zhang2024autocoderover} equips LLM agents with dedicated code-search APIs, supporting iterative retrieval and localization of code fragments related to software bugs. Building on this, \textsc{SpecRover}~\cite{ruan2024specrover} extends \textsc{AutoCodeRover} by focusing on specification inference, generating function summaries, and offering targeted feedback at key points in the agent's workflow. \textsc{Agentless}~\cite{xia2025agentless} demonstrates that even simple LLM agents can address real-world bugs without extensive toolchains or complex modeling of environment behavior. 
%In addition to these leading frameworks, a variety of other agent-based approaches are available in both open-source~\cite{wei2026swe, ouyang2025repograph, bouzenia2025repairagent} and commercial solutions~\cite{wang2025openhands}.

\section{Threats to Validity}
\label{sec:threats-to-validity}

Similar to prior techniques, \approach comes with some limitations and threats to validity. In this section, we discuss how we mitigated various threats.

\noindent\textbf{Internal Validity.}
A major internal threat is that we run each experiment once. As LLMs are non-deterministic, repeated runs may yield different individual test outcomes. However, given the scale of our evaluation ($118$ projects), aggregate metrics are unlikely to change significantly.

\noindent\textbf{External Validity.}
The primary external threat concerns generalizability. To mitigate this, \approach is designed to be PL-agnostic, requiring minimal engineering effort to extend to new PL pairs. Our initial implementation supports six PLs across four PL pairs. A secondary threat is data contamination, as our benchmark programs were likely included in Claude's pre-training data, potentially inflating apparent performance.

\noindent\textbf{Construct Validity.}
To minimize construct validity threats, \approach is built upon well-vetted, widely adopted tools, including Tree-sitter and Claude Code.

\section{Conclusion}
\label{sec:conclusion}

In this work, we introduced \approach, a \emph{language-agnostic repository-level} code translation and validation framework that integrates four specialized LLM agents to achieve high-quality translations validated by both developer-written and agent-generated tests. \approach combines the power of LLMs with reliable static analysis tools to translate projects across four different language pairs and six distinct PLs. To the best of our knowledge, \approach is the first approach that can effectively translate and validate code at the repository level across multiple PLs.

\section{Data Availability Statement}
\label{sec:data-availability}
The artifacts of \approach are publicly available~\cite{website}.

\begin{acks}
This work is supported by NSF CCF-2238045 grant. We thank the anonymous reviewers for their comments, which helped make this work stronger. We also thank Professor Elsa Gunter for her support throughout this project. 
%partially funding one of the authors during this research.
\end{acks}

\bibliographystyle{ACM-Reference-Format}
\bibliography{bibliography}

\end{document}